\documentclass[12pt]{article}

\usepackage{graphicx}
\usepackage{amssymb}
\usepackage[english]{babel}
\usepackage{amsmath,amssymb,setspace,fancyhdr,geometry,url, enumerate}
\usepackage{lscape,amsthm}
\usepackage{amsfonts}
\usepackage{mathtools}
\usepackage{mathrsfs}
\usepackage{algorithm}
\usepackage{graphicx,caption, subcaption, csquotes}
\usepackage{hyperref}
\usepackage{color}
\usepackage{verbatim}
\usepackage{breqn}

\usepackage[mathlines]{lineno}

\usepackage[default]{jasa_harvard}    

\DeclareMathAlphabet{\mathpzc}{OT1}{pzc}{m}{it}

\makeatletter
\def\BState{\State\hskip-\ALG@thistlm}
\makeatother

\begin{document}

\bibliographystyle{ECA_jasa}

\title{A Spatially-Varying Stochastic Differential Equation Model for Animal Movement}
\author{James C Russell, Ephraim M Hanks, Murali Haran and David Hughes}
\maketitle

\begin{abstract}
    \hspace{0.1cm} Animal movement exhibits complex behavior which can be influenced by unobserved environmental conditions. We propose a model which allows for a spatially-varying movement rate and spatially-varying drift through a semiparametric potential surface and a separate motility surface. These surfaces are embedded in a stochastic differential equation framework which allows for complex animal movement patterns in space. The resulting model is used to analyze the spatially-varying behavior of ants to provide insight into the spatial structure of ant movement in the nest.
\end{abstract}



\section{Introduction}\label{intro}

 Studying the movement of animals allows scientists to address fundamental questions in ecology and epidemiology. It can be used to understand how animals are impacted by their environment; for example \citeasnoun{gibert2016crossing} studied the impact of temperature on animal movement to understand potential impacts of climate change  and \citeasnoun{watkins2013evaluating} studied the impact of novel environments on small fish in a simulation-based analysis. Animal movement can give insight into the impact of external stimuli, as studied by \citeasnoun{dodge2013environmental} and \citeasnoun{thiebault2013splitting}. Further, animal movement is important in understanding the collective behavior of animal societies \cite{watkins2013evaluating}. Understanding these factors can benefit animal conservation \cite{killeen2014habitat} and can increase understanding of the impact of migration on disease dynamics \cite{altizer2011animal}.

Improvements in technology have allowed scientists to observe animal movement at a fine temporal resolution over long periods of time \cite{toledo2014lightweight}. This fine scale observation enables researchers to observe more complete individual paths, revealing the animal's behavioral patterns in more detail and allowing for the fitting of more complex statistical models \cite{avgar2015space,bestley2015taking}. Since animal movement data inherently consist of observations through time, it is intuitive to describe movement using differential equations which are often used to describe dynamic systems. As any deterministic differential equation is unlikely to capture the movement of a single animal, we propose a stochastic differential equation (SDE) approach to modeling animal movement. Similar approaches based on stochastic differential equations have been utilized several times to model animal movement \cite{blackwell1997random,preisler2004,Johnson2008,preisler2013}.

In this article, we will focus on the analysis of the movement of ants in a four chambered nest. Ants provide an ideal system for studying animal movement, as an entire population (colony) can be observed and tracked \cite{mersch2013tracking}, something rarely possible outside of laboratory systems. If the entire population had not been observed, like with any model for animal movement, conclusions should be limited to reflect the fact that the movement behavior of an animal directly impacts whether it is observed. For this same reason, ants provide an ideal system for studying the spread of information or infectious disease in societies. Despite the fact that they live in close proximity to one another, researchers have hypothesized a `collective immunity' where collective behavior helps to prevent the spread of disease \cite{quevillon2015social,cremer2007social}. Proposed mechanisms which result in this `collective immunity' include the spatial and social segregation of ants in the nest which minimizes the number of foraging ants which come into contact with the queen. Ant behavior, however, can be challenging to model due to this social and spatial segregation within the nests \cite{quevillon2015social}. In this paper, we propose a spatially-varying  SDE model to capture ant movement behavior within the nest. Analysis of the ant movement indicates a tendency to move more quickly in the central chambers of the nest with ants utilizing the central chambers primarily as corridors for commuting. Simulation of new ants entering the nest highlights that the time needed to pass through sections of the nest varies based on spatial location.

 \citeasnoun{Johnson2008} uses a continuous time correlated random walk model based on integrated Brownian motion to model directional persistence in movement. Here, directional persistence refers to the tendency of animals to continue moving in the same direction and at a similar pace at nearby time points; this can, for instance, induce autocorrelation. However \citeasnoun{Johnson2008} does not consider the case in which the behavior of an animal is dependent on its position. \citeasnoun{quevillon2015social} analyzes the movement of ants in a nest using the continuous time discrete space Markov chain model of \citeasnoun{hanks2015continuous}. The discrete time specification used in \citeasnoun{quevillon2015social} and \citeasnoun{hanks2015continuous} requires the discretization of spatial location using a grid, and analyzing movement by modeling the amount of time spent in each grid cell, and the transition probabilities between cells. Their results reveal spatially-varying movement behavior near the queen.

 Stochastic differential equations have also been used for the movement of objects on a sphere \cite{brillinger2012particle}. This is essential for modeling movement of animals such as elephant seals \cite{brillinger1998elephant} over large distances of the globe. \citeasnoun{brillinger2001modelling} and \citeasnoun{preisler2004} propose an SDE based model for movement on a potential surface which captures spatially-varying drift in movement patterns across space. Potential functions are defined as functions of spatial location, and the negative gradient of this function determine the directional tendencies of animal's movement at a specified location. These potential surface methods are also described in \citeasnoun{brillinger2002} and \citeasnoun{brillinger2012}. Potential surfaces have been used to analyze the movement of monk seals \cite{brillinger2008three} and even the flow of play in soccer \cite{brillinger2007potential}. Potential functions have also been used to model constrained movement \cite{brillinger2003simulating} and pairwise interactions in the movement of particles \cite{brillinger2011modelling}.

 Existing potential surface methods allow for flexible modeling of directional bias in movement(directional force acting on an individual) but cannot easily model variation in the absolute speed of movement when there is no consistent directional bias. This is important in ant systems, as ant nests typically contain ``corridors" which connect important chambers in the nest. Ant movement through these corridors shows high velocity and directional persistence (correlated random walk movement), but little directional bias in these specific regions of the nest as some ants are moving in one direction while other ants are moving in the other. Existing models allow for directional persistence or spatially-varying drift. For example, \citeasnoun{preisler1995autoregressive} analyzes the movement of beetles, and captures directional persistence using an autoregressive model for the turning angles of individuals. We propose an SDE approach to model movement behavior with all three features: (1) correlated movement patterns, (2) spatially-varying drift through a potential surface and (3) spatial variation in movement rates through the inclusion of a spatially-varying motility surface. We define a motility surface as a function of spatial location that determines the average overall movement rate at a specified spatial location. Our approach differs from that utilized by \citeasnoun{quevillon2015social} in that we model movement in continuous space rather than movement between discretized grid cells.

 The remainder of the paper is organized as follows. In Section \ref{data}, we introduce the carpenter ant system in detail. Next, in Section \ref{model} we discuss our proposed model for the ant movement data which incorporates autocorrelation, spatially-varying drift, and spatially-varying absolute movement rate. In Section \ref{inference} we describe a discrete approximation of the model and our Bayesian inferential approach. The results of the application of our approach to ant movement data are presented in Section \ref{antresults}. We conclude in Section \ref{discussion} with a discussion and potential directions for future work.

\section{Carpenter Ant Movement}\label{data}

We begin with a description of the ant movement data and describe an exploratory data analysis that motivates our modeling approach. We analyze the movement of the ants in a custom-constructed nest. We used the common black carpenter ant \textit{Camponotus pennsylvanicus} which nests in wood in temperate forests in the Eastern USA. We collected colonies between May and June 2015. The ants were placed in a nest structure consisting of four distinct chambers. Each chamber is divided into two sections by an internal barrier, creating a small passageway $12$mm across between the upper and lower halves of the chamber. Each of the four chambers measures $65$mm by $40$mm resulting in a total nest size of $65$mm by $160$mm. Each doorway between chambers is $6$mm across. There is an exit from the nest in chamber IV, leading to an area with food and water. The queen resides primarily in chamber I, far from the nest exit.

\begin{figure}
\centering
\begin{subfigure}{.99\textwidth}
  \centering
  \includegraphics[clip, trim=0.25cm 0cm 0.05cm 5.25cm, width=0.6\textwidth]{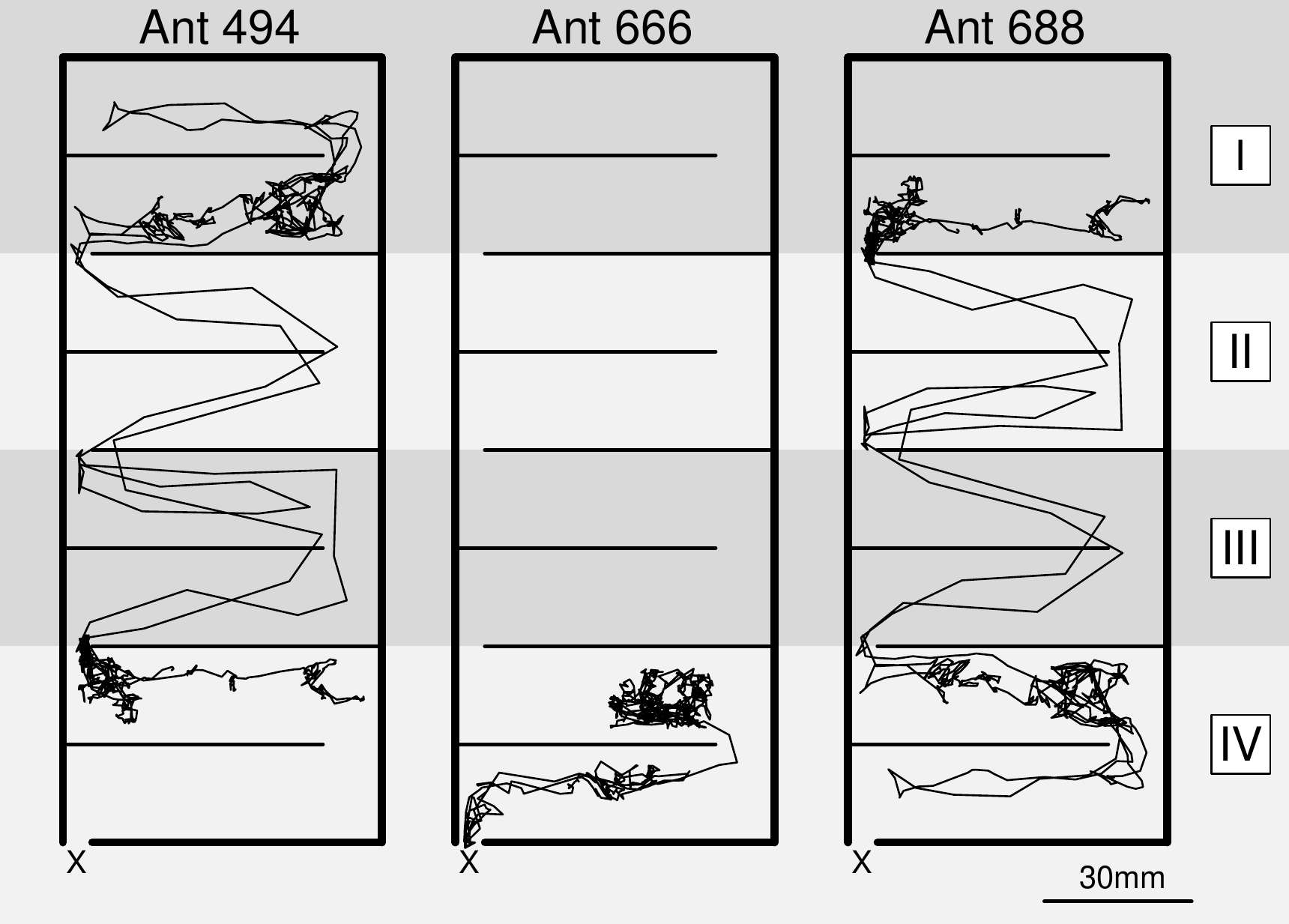}
\end{subfigure}%
\caption{ Movement paths of three sample ants.}
\label{fig:AntNest}
\end{figure}

A plot containing three examples of individual ant movement paths, using linear interpolation between observed locations at each second, is given in Figure \ref{fig:AntNest}. The nest exit is marked with an ``X" below chamber IV in each plot. The movement paths indicate that ant behavior is different within different chambers of the nest, with faster, more directed movement happening in chambers II and III. This type of spatially-varying behavior would be difficult to capture using the potential surface SDE approach (e.g \citeasnoun{brillinger2002} and \citeasnoun{preisler2004}) as the ants move quickly in some regions of the nest but do not show consistent drift.

The data consist of the 2-dimensional location coordinates $\left( x_{t_i,j}, y_{t_i,j} \right) $, a time index $t_i$, which ranges from 1 to 3600 for the one hour observation period. In this case the observations are made at evenly-spaced intervals, however this is often not the case in movement data. Additionally, there is a unique ant identifier $j$ for each of $32$ ants. One camera was positioned over each chamber of the nest. Ants were filmed under infrared lighting with GoPro Hero3 and Hero3+ cameras with modified infrared filters (RageCams, Sparta, MI) to simulate natural lighting conditions. Each ant was individually tagged with a unique identifier allowing a human observer to record their position. Observations were recorded at every second by clicking on the location of the ant using a custom software package. Measurement error is small and there are no missing observations as the recorded videos allow for careful location selection and, if necessary, review of the movements of each individual. Locations were recorded for the entire hour for all ants that enter chamber IV at any time during the observation window. The dimensions of the nest were recorded by clicking on predetermined corners in the nest at the beginning of the observation period. Each camera records a separate section of the nest, so that there are four separate sets of observation for each ant. The observations in each of the four chambers can then be combined, using a common time index, to represent movement across the four chambers.

Several challenges arise in combining the data from the four cameras. At some time-points (0.02\% of observations) an ant is positioned at the door between between two chambers and is observed by two cameras in two chambers at the same time. This could be due to different segments of the ants being visible in different chambers at the same time-point. Further complications arise when the ants are not observed for a span of time. This can happen when the ant has exited the nest structure (5.46\% of observations) or when the ant is situated in chamber doorways (1.08\% of observations). When the ant has exited the nest structure no cameras capture its movement. In what follows, we assume that when the ant re-enters the nest its movement is independent from prior in-nest movements. If the ant is between chambers and not visible on any camera, the ant's locations are linearly interpolated from the observations before and after the ant is in the entryway. 

\begin{figure}
\centering
\begin{subfigure}{.5\textwidth}
  \centering
  \includegraphics[width=.95\linewidth]{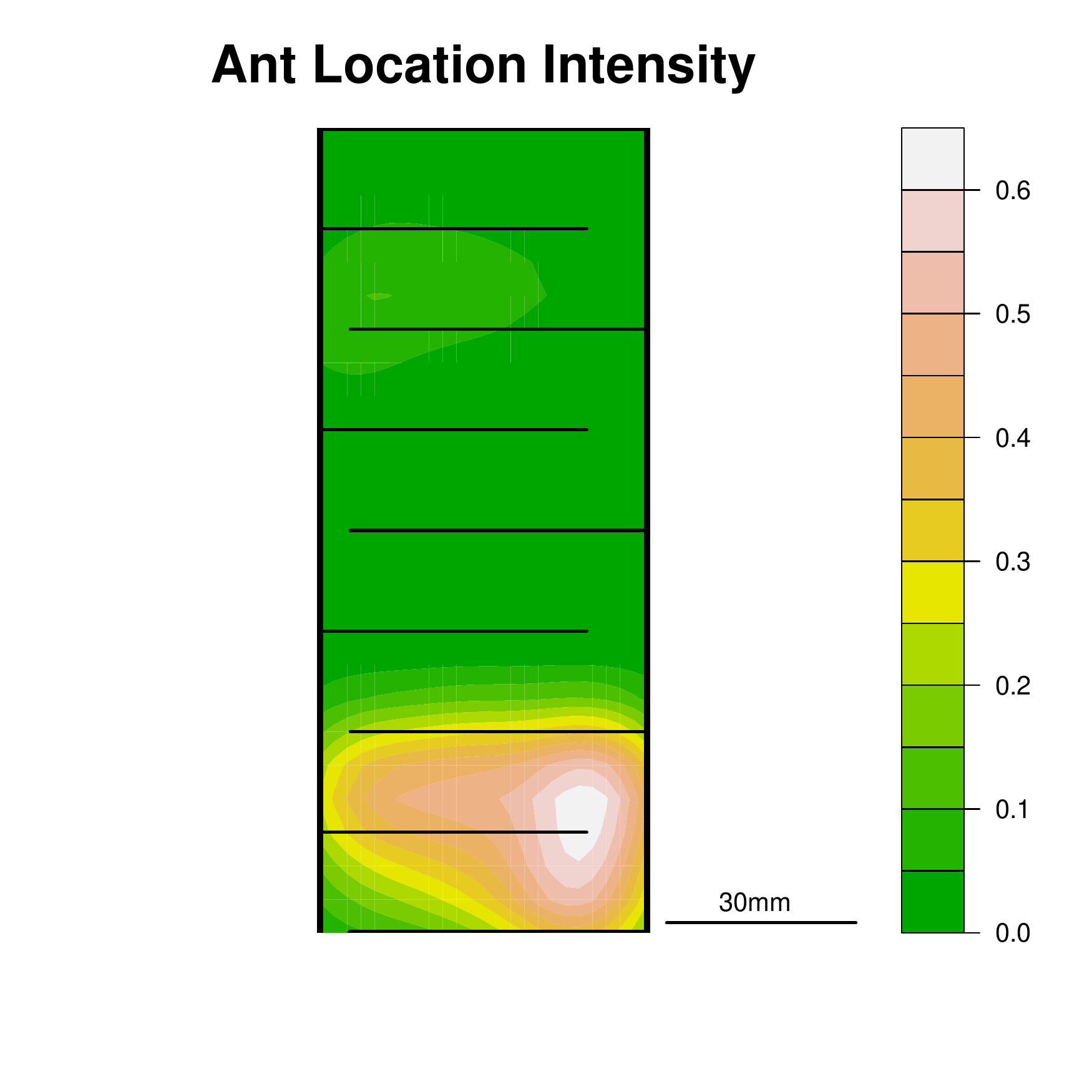}
  \caption{Ant Location Intensity Surface}
  \label{fig:EDA1}
\end{subfigure}%
\begin{subfigure}{.5\textwidth}
  \centering
  \includegraphics[width=.95\linewidth]{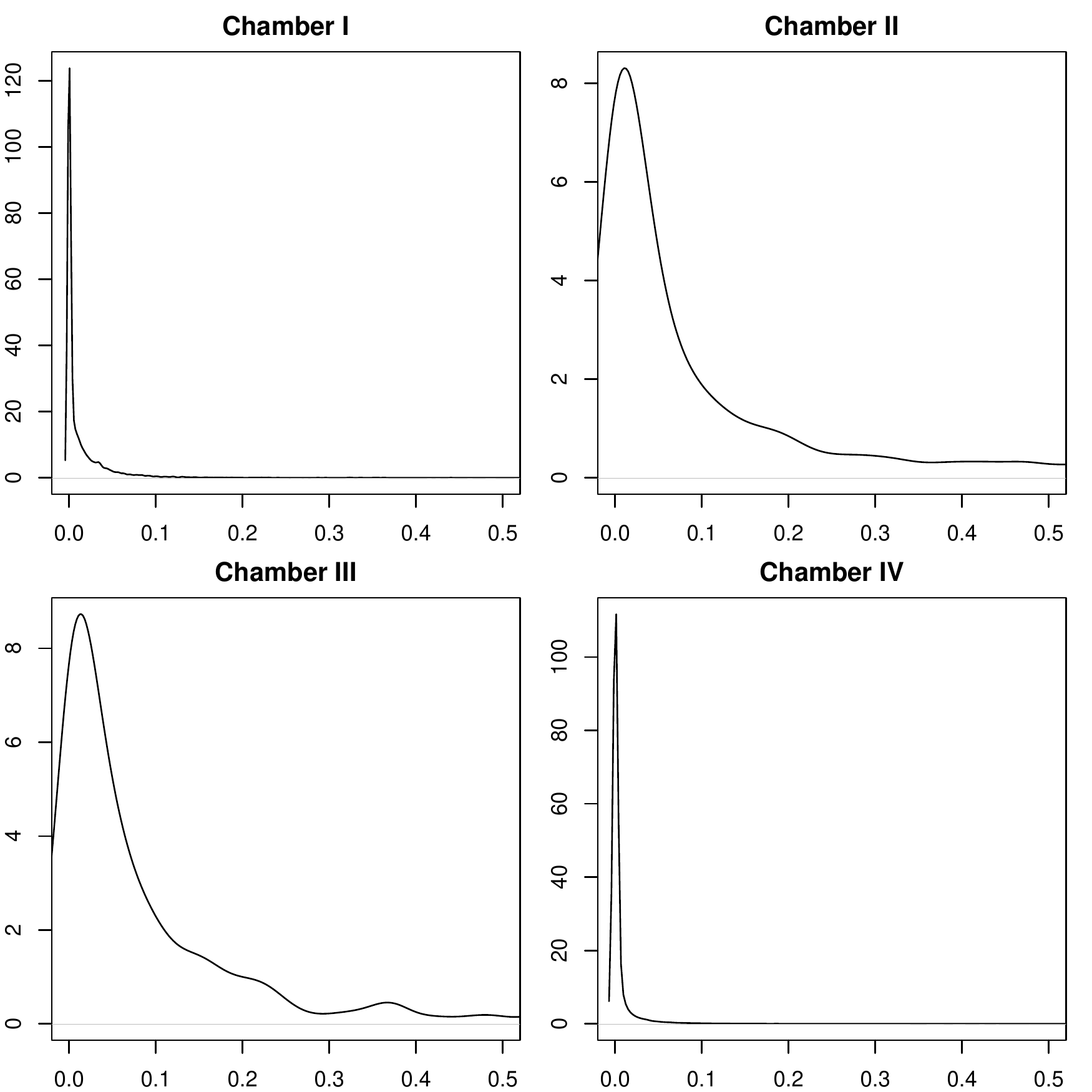}
  \caption{Empirical Velocities in Each Chamber}
  \label{fig:EDA2}
\end{subfigure}
\caption{Exploratory Data Analysis for carpenter ant movement.}
\label{fig:EDA}
\end{figure}

To explore the spatial movement behavior of the ants, a kernel density estimate for all observed ant locations and kernel density estimates of all observed empirical velocities in each of the nest chambers are plotted in Figure \ref{fig:EDA}. In both cases the kernel density estimates are calculated using the density function of the base stats package in R \cite{ihaka1996r} using a Gaussian kernel and the default bandwidth. From Figure \ref{fig:EDA1}, we can see that the ants spend most of their time on the right-hand side of chamber IV, this indicates that there are certain preferred regions of the nest that ants tend to move toward, hinting at spatially-varying drift. Figure \ref{fig:EDA2} indicates that the average velocities in chambers I and IV tend to be less than the average velocities in chambers II and III. These exploratory results indicate that both the drift and the absolute ant velocity may vary spatially and suggest that the center chambers are primarily used for higher velocity transit. To explore the temporal autocorrelation in movement, an autoregressive model of order 1 was fit to the empirical velocities of a randomly selected ant. The resulting estimated autoregressive parameter, $0.81$ (standard error $0.015$) indicates the need to model directional persistence. Our exploratory data analysis suggests that an appropriate model for these data should therefore include temporal autocorrelation, spatially-varying drift, and a spatially-varying movement rate. We develop a model which allows for this type of behavior in the Section \ref{model}.

\section{Spatially-Varying SDE Model}\label{model}

In this section, we propose an SDE model for animal movement that captures
\begin{enumerate}
\item directional persistence via a continuous time correlated random walk (CTCRW),
\item spatial variation in drift through a potential surface,
\item spatial variation in overall movement rate using a motility surface.
\end{enumerate}
These three features have not been jointly utilized in any previous models. In previous work, potential surfaces have been used ignoring temporal autocorrelation to simplify the model and for computational feasibility \cite{preisler2004,brillinger2012}. We extend these methods by modelling directional persistence through a CTCRW, and expand the model to allow for variability in movement rate. This model will be used to analyze within-nest ant movement, where at time t, an individual's location at time t is denoted $\left(x(t), y(t)\right)$ and its velocity is denoted $\left(v_x(t), v_y(t)\right)$. We will begin by describing the movement of one individual, and in Section \ref{inference}, we will generalize for multiple independent ants.

\subsection{Continuous Time Correlated Random Walk}\label{CTCRW1}

 Animal movement is often autocorrelated through time, resulting in smooth movement paths. The CTCRW model presented by \citeasnoun{Johnson2008} provides one method for modeling dependence through time. The CTCRW model specifies an Ohrstein-Uhlenbeck (O-U) process on an animal's velocity with directional drift $\mu$. The O-U process is defined as a stochastic process that is stationary (the joint density remains the same for uniform shifts in time), Gaussian, Markovian and has continuous paths \cite{klebaner2005introduction}. Alternatively, the O-U process can be derived using linear stochastic differential equations \cite{gardiner1986handbook}.

We consider a process that describes the movement of a particle in two dimensions. The movement in each dimension is assumed to follow an independent CTCRW. This is a reasonable assumption for animal movement since positive correlation in the x-dimension and y-dimension would result in bias toward movement in the North-East or South-West directions. The CTCRW model to describe the velocities $v_x(t)$,$v_y(t)$  and the locations $x(t)$,$y(t)$  is defined as
\begin{linenomath*}
\begin{align*}
dv_x(t) &= \beta \left( \mu_x - v_x(t)\right) dt + \sigma dW_{v_y}(t) \\
dv_y(t) &= \beta \left( \mu_y - v_y(t)\right) dt + \sigma dW_{v_x}(t) \\
dx(t) &= v_x(t)dt + \kappa dW_{x}(t)\\
dy(t) &= v_y(t)dt + \kappa dW_{y}(t).
\end{align*}
\end{linenomath*}
In the velocity equations, $\mu_x$ represents the mean drift in the x direction, $\mu_y$ represents the mean drift in the y direction, $\beta$ controls the autocorrelation in movement, $\sigma$ is related to the variability in velocity, and $W_{v_x}(t)$, $W_{v_y}(t)$ represent independent Brownian motion processes with unit variance. In the equation for location, $\kappa$ is related to the additional variability in location (that is measurement error, which relates to the error in ``clicking location" when recording the locations of the ant from the recorded videos) and $W_{x}(t)$, $W_{y}(t)$  are again Brownian motion with unit variance. The case where $\kappa=0$ gives intuition on the position of the object over time
\begin{linenomath*}
\begin{align}\label{eq:simplelocation}
x_t=x_0 + \int_{0}^t v_x(s)ds.
\end{align}
\end{linenomath*}
The CTCRW process results in a model where the conditional mean of the discretized velocity is a weighted average of the previous velocity and the directional drift term $\mu$. Several R packages have been developed due to the popularity of the CTCRW model for animal movement. For example, \citeasnoun{albertsen2015fast} introduces the Template Model Builder (TMB) package which incorporates non-gaussian error in a computationally efficient manner through Laplace approximation. Other examples of R packages to fit the CTCRW model include crawl \cite{johnson2013crawl} and bsam \cite{jonsencontributions}. Examples of the use of this model include \citeasnoun{baylis2015disentangling}, which analyzes the decline in population of southern sea lions, \citeasnoun{northrup2015quantifying}, which studies the impact of habitat loss on the movement of mule deer, and \citeasnoun{rode2015increased}, which analyzes the movement of polar bears in reaction to the decrease in sea ice habitat. The CTCRW model, however, assumes that the movement behavior is homogeneous in space. This assumption can be relaxed by allowing the drift term for the individual's velocity to be a function of the individual's location $\left(x(t), y(t)\right)$. Doing so unifies the CTCRW model with the potential function models of \citeasnoun{brillinger2002} and \citeasnoun{preisler2004}.

\subsection{Potential Surface}\label{potsurf}

Potential surfaces are described in relation to animal movement modeling by \citeasnoun{brillinger2002} and \citeasnoun{preisler2013}. There are several examples of analyses using potential surfaces in movement modeling, including \citeasnoun{preisler2004}, which studies the influence of roads on the movement of elk, and \citeasnoun{brillinger2012}, which analyzes the movement of elk in a fenced-in experimental forest. In a potential function approach, the individuals, or ants in our case, are considered to be moving around on a surface with regions of attraction and repulsion.

First consider a potential surface, $H(x,y)$, that is only a function of the current two-dimensional location of the object $(x,y)$. This function can be thought of as a topological surface where objects are drawn to lower regions similar to a marble moving on a curved surface. The mean direction of movement is therefore down the slope of the surface, or equivalently in the direction of the negative gradient of the potential surface. Using this analogy, the expectations of the $x$ and $y$ components of velocity, denoted $v_x(t)$ and $v_y(t)$ respectively, of an individual can each be calculated by taking the negative of the $x$ and $y$ derivatives of $H(x,y)$ respectively:
\begin{linenomath*}
\begin{align*}
E\left(v_x(t)\right) &= - \frac{dH(x(t),y(t))}{dx}\\
E\left(v_y(t)\right) &= - \frac{dH(x(t),y(t))}{dy}
\end{align*}
\end{linenomath*}
This spatially-varying drift in movement is incorporated in the CTCRW model from Section \ref{CTCRW1} by defining the mean drift $\mu$ as the negative gradient of $H(x,y)$. The result is a system of stochastic differential equations for correlated velocity with spatially-varying movement bias defined by the potential surface $H(x,y)$
\begin{linenomath*}
\begin{equation}\label{eq:SDE1}
\begin{aligned}
dv_x(t) &= \beta \left( - \frac{dH\left(x(t),y(t)\right)}{dx}  - v_x(t)\right) dt + \sigma dW_{v_x}(t)\\
dv_y(t) &= \beta \left( - \frac{dH\left(x(t),y(t)\right)}{dy}  - v_y(t)\right) dt + \sigma dW_{v_y}(t)
\end{aligned}
\end{equation}
\end{linenomath*}
The SDE model for velocity \eqref{eq:SDE1} jointly models spatially-varying drift through the potential function $H(x,y)$ and temporal autocorrelation through the finite autoregressive parameter $\beta$. In previous work (e.g. \citeasnoun{brillinger2002} and \citeasnoun{preisler2004}), only the over-damped case is considered, where $\beta \rightarrow \infty$. Taking the limit as $\beta \rightarrow \infty$ eliminates directional persistence in movement, as it puts all of the importance on the gradient of the potential surface, and eliminates the dependence on the animal's current velocity in \ref{eq:SDE1}. This simplifies estimation of the potential surface, but fails to model the temporal autocorrelation which is normally present in animal movement.

We model the potential surface using B-spline basis functions \cite{de1978practical}. Parametric models of the potential surface are also possible \cite{quevillon2015social,hanks2015continuous} but our goal in this analysis is to flexibly model spatially-varying behavior within the nest. We thus assume the potential surface is given by
\begin{linenomath*}
 \begin{align}\label{eq:PotentialSurface}
 H( x, y) = \sum_{k,l} \gamma_{kl} \phi_k^{(M)} (x) \psi_l^{(M)} (y)
 \end{align}
\end{linenomath*}
 where $\phi_k^{(M)} (\mathpzc{x})$ and $\psi_l^{(M)}  (\mathpzc{y})$ are B-spline basis functions of order $M$, with $K$ basis functions in the $x$-direction and $L$ basis functions in the $y$-direction so that $k \in \{1,...,K\}$ and $l \in \{1,...,L\}$. We chose to set $M=4$ so that the potential surface $H(x,y)$ has two continuous derivatives. As a result, the gradient of the potential surface \eqref{eq:PotentialSurface} also has a continuous derivative, meaning it is relatively smooth. The potential surface $H(x,y)$ only impacts movement through its gradient. Thus only contrasts of the B-spline coefficients $\{\gamma_{kl}\}$ are identifiable and we elect to subject them to the constraint $\sum_{k,l} \gamma_{kl}=0$. Rather than using tuning to select the number of knots directly, the number of basis functions is set to $560$, selected to keep the scale in the dimensions ($65$mm by $160$mm) approximately equal (K=16, L=35) and penalized \cite{eilers1996} by using a zero mean multivariate normal prior on the coefficients $\{\gamma_{kl}\}$. Details of this prior will be given in Section \ref{inference}.

The movement of the ants in our analysis is restricted by the locations of the walls in the ant chambers (Figure \ref{fig:AntNest}). Various models that account for restricted animal movement have been proposed \cite{brillinger2003simulating,brost2015animal}. In this case, we restrict ant movement near walls  by augmenting the spatially smooth potential function \eqref{eq:PotentialSurface} with an additive exponential potential $R(x,y)$
\begin{linenomath*}
\begin{align*}
R(x,y, \boldsymbol{r}, r_1)= &\text{exp}\{-r_1 (x-r^x_l))\} + \text{exp}\{r_1(x-r^x_u)\} +\\
 &\text{exp}\{-r_1(y-r^y_l)\} + \text{exp}\{r_1(y-r^y_u)\}
\end{align*}
\end{linenomath*}
where $\boldsymbol{r}=(r^x_u, r^x_l, r^y_u, r^y_l)$; $r^x_u$ and $r^x_l$ represent the upper and lower wall boundaries in the x-dimension; and $r^y_u$ and $r^y_l$ represent the upper and lower wall boundaries in the y-dimension. The parameter $r_1$ controls the rate of decay of the wall repulsion function. Other formulations for this basis are possible, but estimation using more flexible wall basis functions may require observations with a higher temporal frequency near the boundaries. The combined potential surface we consider is
\begin{linenomath*}
\begin{align}\label{eq:CombinedPotential}
 H( x, y) = \sum_{k,l} \gamma_{kl} \phi_k^{(M)} (x) \psi_l^{(M)} (y) + R(x,y, \boldsymbol{r}, r_1).
 \end{align}
\end{linenomath*}

\subsection{Motility Surface}\label{fricsurf}

In previous  studies of the movement of ants in a nest, researchers have found that ants tend to move at different speeds in different areas of their nest \cite{quevillon2015social}. To incorporate this behavior in our SDE model, we propose a spatially-varying motility surface $M(x(t),y(t))$ which scales the overall rate of movement in different parts of the nest. The motility surface scales the absolute movement rate in different spatial locations, allowing for flexible modelling of animal movement in which velocities can depend on the individual's environment. This results in the following stochastic differential equation model, where the location equation from the CTCRW model of \citeasnoun{Johnson2008} in each dimension has been adjusted to account for the spatially-varying motility
\begin{linenomath*}
\begin{equation}\label{eq:SDE2}
\begin{aligned}
dx(t) &= M(x(t),y(t))v_x(t)dt + \kappa dW_{x}(t)\\
dy(t) &= M(x(t),y(t))v_y(t)dt + \kappa dW_{y}(t).
\end{aligned}
\end{equation}
\end{linenomath*}
When $\kappa=0$, we get the physical interpretation of velocity scaled by the motility surface $x(t) = \int_0^t M(x(s),y(s))v_x(s)ds$, and when $M(x,y)=1$, we get back \eqref{eq:simplelocation}.

For an example of movement behavior induced by the motility surface, consider a habitat separated into two distinct regions, one in which ants move fast, and another in which ants move more slowly. If $M(\cdot)=2$ in the fast region, and $M(\cdot)=1$ in the slow region, the ants will move approximately twice as fast in the fast region as they will in the slow region. Equations \eqref{eq:SDE1} and \eqref{eq:SDE2} together define a continuous time 2-dimensional stochastic movement process. Note that the $v_x$ and $v_y$ can no longer be interpreted directly as velocity, but instead are a scaled version of the animal's movement rate; at $(x(t),y(t))$ the animal's mean movement rate vector ignoring autocorrelation is $\left( M(x,y) \cdot v_x(t) ,  M(x,y) \cdot v_y(t) \right)$.

For the motility surface $M(x(t),y(t))$, we again utilize penalized B-spline basis functions.
\begin{linenomath*}
 \begin{align}\label{eq:FrictionSurface}
 M( x, y) = \sum_{q,r} \alpha_{qr} \zeta_q^{(P)} (x) \xi_r^{(P)} (y)
 \end{align}
\end{linenomath*}
 where $\zeta_q^{(N)} (\mathpzc{x})$ and $\xi_r^{(N)}  (\mathpzc{y})$ are b-spline basis functions of order $P$, with $Q$ basis functions in the $x$-direction and $R$ basis functions in the $y$-direction so that $q \in \{1,...,Q\}$ and $r \in \{1,...,R\}$. To facilitate computation, we set $P=4$, $Q=K$, and $R=L$ so that the motility surface and the potential surface share the same basis functions.

 \subsection{Identifiability}\label{identify}

  Without imposing constraints, there are multiple (infinitely many) combinations of parameters which will yield an identical formula for the model defined by equations \eqref{eq:SDE1} and \eqref{eq:SDE2}. First, the potential surface $H(x,y)$ only enters the model through its gradient. Therefore adding some constant $c_1$ to the surface results in an equivalent model ($H^*(x,y)=H(x,y)+c_1$). As discussed in Section \ref{potsurf}, the constant $c_1$ is fixed to $0$ by constraining the sum of the basis coefficients $\{\gamma_{kl}\}$ to equal $0$.

 Second, the motility surface and velocity terms are only identifiable up to a multiplicative constant $c_2$. Multiplying the motility surface by $c_2$ and the velocities by $\frac{1}{c_2}$ yields an equivalent model
 \begin{linenomath*}
\begin{equation*}
\begin{aligned}
\frac{1}{c_2} dv_x(t) &= \beta \left( - \frac{1}{c_2} \frac{dH\left(x(t),y(t)\right)}{dx}  - \frac{1}{c_2} v_x(t)\right) dt + \frac{1}{c_2} \sigma dW_{v_x}(t)\\
\frac{1}{c_2} dv_y(t) &= \beta \left( - \frac{1}{c_2} \frac{dH\left(x(t),y(t)\right)}{dy}  - \frac{1}{c_2} v_y(t)\right) dt + \frac{1}{c_2} \sigma dW_{v_y}(t)
\end{aligned}
\end{equation*}
\end{linenomath*}
\begin{linenomath*}
\begin{equation}\label{eq:identify}
\begin{aligned}
dx(t) &= c_2 M(x(t),y(t)) \frac{1}{c_2} v_x(t)dt + \kappa dW_{x}(t)\\
dy(t) &= c_2 M(x(t),y(t)) \frac{1}{c_2} v_y(t)dt + \kappa dW_{y}(t).
\end{aligned}
\end{equation}
\end{linenomath*}
To obtain identifiability, the model is constrained by setting $c_2=\sigma^2$, or equivalently, fixing $\sigma^2=1$.

\section{Inference}\label{inference}

  It is impossible to analytically solve the non-linear SDEs given by Equations \eqref{eq:SDE1} and \eqref{eq:SDE2} so instead we consider approximate solutions using the Euler-Maruyama method \cite{kloeden1992numerical} which is based on a first order Taylor expansion. We approximate the continuous time process \eqref{eq:SDE1} and \eqref{eq:SDE2} at a set of observed (or simulated) discrete-times $t_i$, $i=1,2,...N_{sim}$, where $N_{sim}$ is the number of points in the path (observed or simulated). Higher order Euler-Maruyama approximations could be used to improve accuracy, and these higher order methods may be interesting directions for future research, both in the context of simulation and inference. Higher order approximations are not necessary in this case due to the high temporal frequency of discrete observations, every second for the ants. An analysis of the impact of the time discretization in the Euler-Maruyama method is presented in the appendix. The Euler-Maruyama method results in the following discrete approximations for Equations \eqref{eq:SDE1} and \eqref{eq:SDE2}
  \begin{linenomath*}
 \begin{equation}\label{eq:EM}
 \begin{aligned}
 v^{x}_{t_{i+1}} &= v^{x}_{t_{i}} + \beta \left( - \nabla H\left(x_{t_{i}},y_{t_{i}}\right)  - v^{x}_{t_{i}}\right) \Delta_t + \sigma \epsilon_{2x} \\
 v^{y}_{t_{i+1}} &= v^{y}_{t_{i}} + \beta \left( - \nabla H\left(x_{t_{i}},y_{t_{i}}\right)  - v^{y}_{t_{i}}\right) \Delta_t + \sigma \epsilon_{2y}\\
 x_{t_{i+1}} &= x_{t_{i}} + M(x_{t_{i}},y_{t_{i}})v^{x}_{t_{i}}\Delta_t + \kappa \epsilon_{1x} \\
 y_{t_{i+1}} &= y_{t_{i}} + M(x_{t_{i}},y_{t_{i}})v^{y}_{t_{i}}\Delta_t + \kappa \epsilon_{1y}
 \end{aligned}
 \end{equation}
 \end{linenomath*}
 where $\Delta_t$ is the time step of the discretized solver and  $\epsilon_{1x}, \epsilon_{1y}, \epsilon_{2x}, \epsilon_{2y}$ are independent Gaussian random variables with mean $0$ and variance $\Delta_t$. These discrete Euler-Maruyama approximations will be used for inference. This is similar to \citeasnoun{wikle2010general}, which uses a second-order Taylor expansion to approximate nonlinear dynamics in statistical models with differential equations. A 2-dimensional simulation generated using the Euler-Maruyama method is presented in the Appendix to illustrate the range of behavior possible under our SDE model.

    We use Bayesian inference to estimate the model parameters. The Euler-Maruyama approximations of the SDEs are used to provide an approximation for the likelihood function in the posterior distribution. Markov chain Monte Carlo is used to draw samples from this approximate posterior distribution. Vague priors are used for the the location variability parameter $\kappa^2 \sim \mbox{Gamma}(0.001,0.001)$, and for the autocorrelation parameter $f_{\beta}\left( \beta \right) \propto \mbox{exp}\left( \frac{-(\beta-1)^2}{20000} \right)I(\beta>0)$, a normal distribution truncated to be positive. To test sensitivity to prior specification, inference was performed again using priors with twice the variance, and there was no significant difference in the results. For identifiability $\sigma^2$ is fixed at $1$. To induce spatial smoothness in the potential and motility surfaces, a proper CAR covariance structure \cite[Chapter~3]{banerjee2014hierarchical} is used for the coefficients of both of the B-spline expansions in \eqref{eq:PotentialSurface2} and \eqref{eq:FrictionSurface}. The model was also fit with an uncorrelated prior on all of the spatial surface parameters, and the resulting surface estimates were very jagged and indicated overfitting. Thus, for the potential surface coefficients, the prior selected is:
    \begin{linenomath*}
    \begin{align*}
    \boldsymbol{\gamma} \sim N\left(\boldsymbol{0}, \left(\tau_{\gamma}\left( \boldsymbol{D} -\rho_{\gamma} \boldsymbol{Q} \right)  \right)^{-1} \right)
    \end{align*}
    \end{linenomath*}
    Where $(\boldsymbol{D})_{jj} = $ \{number of neighbors of j\}; $(\boldsymbol{D})_{ij} = 0$ if $i\neq j$; $(\boldsymbol{Q})_{jj} = 0$ and $\boldsymbol{Q}_{ij}= I(i,j \mbox{ are neighbors})$. For the smoothness parameter we use the prior $\rho_{\gamma} \sim \mbox{Uniform}(0.01,0.99)$. The parameter $\tau_{\gamma}$ is a tuning parameter that determines the scale of the potential surface. The prior distribution for $\tau_{\gamma}$ is set to be an exponential distribution $\tau_{\gamma} \sim \mbox{exp}(\mu_{\alpha}^2)$ so that the observed data will inform the level of tuning, and the distribution of the potential surface will scale with the level of the motility surface through $\mu_{\alpha}$. As discussed in Section \ref{identify}, only contrasts of the B-spline coefficients $\{\gamma_{kl}\}$ are identifiable, so we utilize the constraint $\sum_{k,l} \gamma_{kl}=0$. This is done through a linear transformation of a set of unconstrained basis function coefficients $\boldsymbol{\tilde{\gamma}}$ \cite[Chapter~12]{gelfand2010handbook}
    \begin{linenomath*}
    \begin{align}\label{eq:PotentialSurface2}
    \boldsymbol{\gamma} = \boldsymbol{\tilde{\gamma}}-\boldsymbol{\Sigma} \boldsymbol{1}^t \left( \boldsymbol{1} \boldsymbol{\Sigma} \boldsymbol{1}^t \right)^{-1}\left( \boldsymbol{1}\boldsymbol{\tilde{\gamma}} - \boldsymbol{0} \right)
    \end{align}
    \end{linenomath*}
    where $\boldsymbol{\Sigma}$ represents the covariance matrix of the unconstrained coefficients.

    Similarly, the coefficients of the B-spline basis functions for the motility surface were assigned a CAR prior
    \begin{linenomath*}
    \begin{align*}
    \boldsymbol{\alpha} \sim N \left(\mu_{\alpha} \boldsymbol{1}, \mu_{\alpha}^2 \left(\tau_{\alpha}\left( \boldsymbol{D} -\rho_{\alpha} \boldsymbol{Q} \right) \right)^{-1} \right).
    \end{align*}
    \end{linenomath*}
    For the motility surface the prior for the smoothing parameter is again set to $\rho_{\alpha} \sim \mbox{Uniform}(0.01,0.99)$. For identifiability, as discussed in Section \ref{identify}, the tuning parameter $\tau_{\alpha}$ is set to $9$ so that $99.8\%$ of the prior mass is positive. The parameter $\mu_{\alpha}$ adds flexibility to the model as it scales the motility surface. The inclusion of $\mu_{\alpha}$ allows for approximating $c_2$ in Equation \eqref{eq:identify}, and therefore, it can be used to approximate $\sigma_2$ if the identifiability constant in Equation \eqref{eq:identify} had instead been fixed at 1. Additionally, $\mu_{\alpha}$ alters the smoothing such that the penalization on variability by the multivariate Gaussian prior in the motility surface coefficients is less in regions of greater relative motility.  A Gaussian prior is used for the scaling parameter $\mu_{\alpha} \sim \mbox{N}\left( 1, 1 \right)$.  A lognormal distribution is used for the prior on the wall repulsion parameter $r_1 \sim \mbox{logNorm}\left( 10, 1 \right)$. The analysis was re-run with sparse priors (the variability of each prior distribution was doubled) and there was no significant changes in the posterior means or credible intervals for all parameters. Full-conditional distributions are available for most parameters in the model, and are presented in the Appendix.

    Block-update MCMC is used to sample from the posterior distributions of model parameters. When available, the updates are drawn from full-conditional distributions (a Gibbs sampler). The coefficients of the potential and motility surfaces, $\boldsymbol{\gamma}$ and $\boldsymbol{\alpha}$, are each updated as a separate block. $10^5$ samples are drawn from the posterior distribution and convergence of the Markov chains is determined by monitoring Monte Carlo standard errors using the batch means procedures \cite{jones2006fixed,flegal2008markov}. The initial $20000$ values of the chain are discarded as burn in, as the initial estimates for the potential surface and motility surface are difficult to select, resulting in inaccurate parameter estimates at the beginning of the chain. Multiple chains with different starting values were run to ensure estimates are robust across initial values.

    Inference here is computationally taxing, as it takes approximately 6 days to generate $10^5$ samples from the posterior distribution on a single core of a 2.7 GHz Intel Xeon Processor with code written in R. The computing time scales at a linear rate with increases in observations. Reducing the computation time needed for inference is one of the primary goals of ongoing research. In this case, the computational complexity of the model  is driven by the large number of latent variables (213,534 velocities $v^{(x)}_{t,j}$ and $v^{(y)}_{t,j}$), and the number of basis functions used for the potential and motility surfaces. Simulations under various settings show that developing a method to accurately approximate latent velocities and reducing the number of basis functions for the surfaces greatly improves computational efficiency. Bayesian implementation allows for straightforward estimation of tuning parameters for the smoothness of our B-spline surfaces, and avoids the additional difficulties of selecting the number of basis functions using other methods such as generalized cross validation. Improving the computational efficiency of inference is an important direction for future work, especially due to the availability of data sets with more individuals and observation over longer time periods.

\section{Results}\label{antresults}

\begin{table}
\begin{center}
  \caption{Ant results}\label{table:antParams}
  \begin{tabular}{| l | c  | c |}
    \hline
    Parameter  & posterior mean & credible interval  \\ \hline
    $\beta$    & $0.872$  & $(0.866, 0.877) $  \\
    $\kappa$ & $0.00133$  & $(0.00131, 0.00134) $  \\
    $r_1$ & $0.078$  & $(0.044, 0.114) $  \\
    \hline
  \end{tabular}\\
    Parameter Estimates for Ant Movement Data.
\end{center}
\end{table}

 Posterior sample means and 95\% credible intervals for the model parameters are presented in Table \ref{table:antParams}. There is significant autocorrelation in the ant's velocity as the credible interval for $\beta$ does not include $1$, but the autocorrelation is not strong. Note that in a continuous time framework (where $\Delta \rightarrow 0$) no autocorrelation corresponds to $\beta \rightarrow \infty$, but in the discrete approximation with $\Delta=1$, no autocorrelation in velocity corresponds to $\beta=1$. This is reasonable for ant movement, as the ant paths are not smooth since the ants tend to change direction suddenly inside the nest. The estimate for $\kappa$ is small, which indicates that there is not much additional location variability after conditioning on latent velocity. Our estimate for $r_1$ indicates that there is a repulsion behavior from the wall, but the size of the effect is relatively small.

\begin{figure}
\centering
\begin{subfigure}{.32\textwidth}
  \centering
  \includegraphics[clip, trim=1.5cm 1.35cm 0.5cm 1.5cm, width=.95\linewidth, page=1]{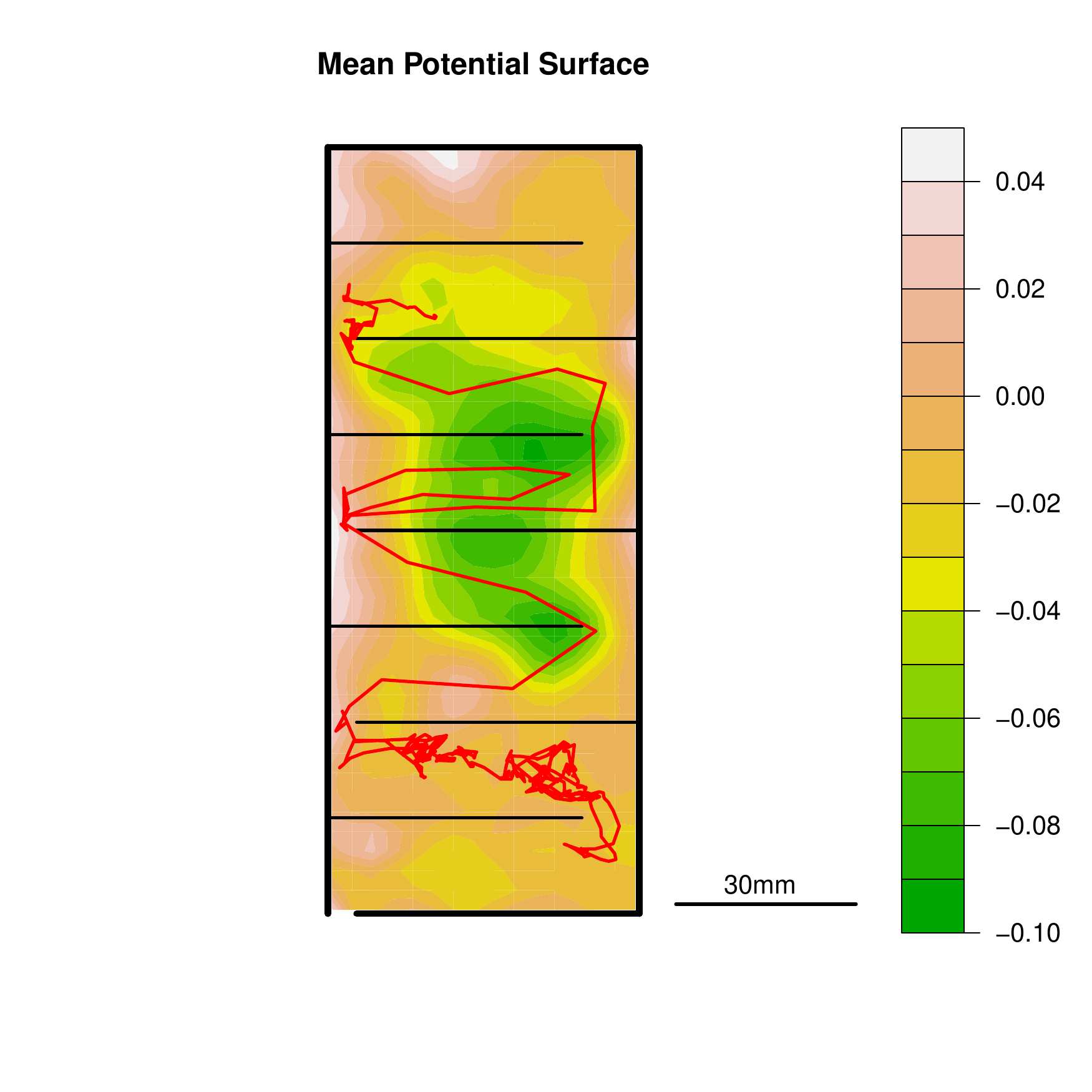}
  \caption{Potential Surface}
\end{subfigure}%
\begin{subfigure}{.32\textwidth}
  \centering
  \includegraphics[clip, trim=1.5cm 1.35cm 0.5cm 1.5cm, width=.95\linewidth, page=1]{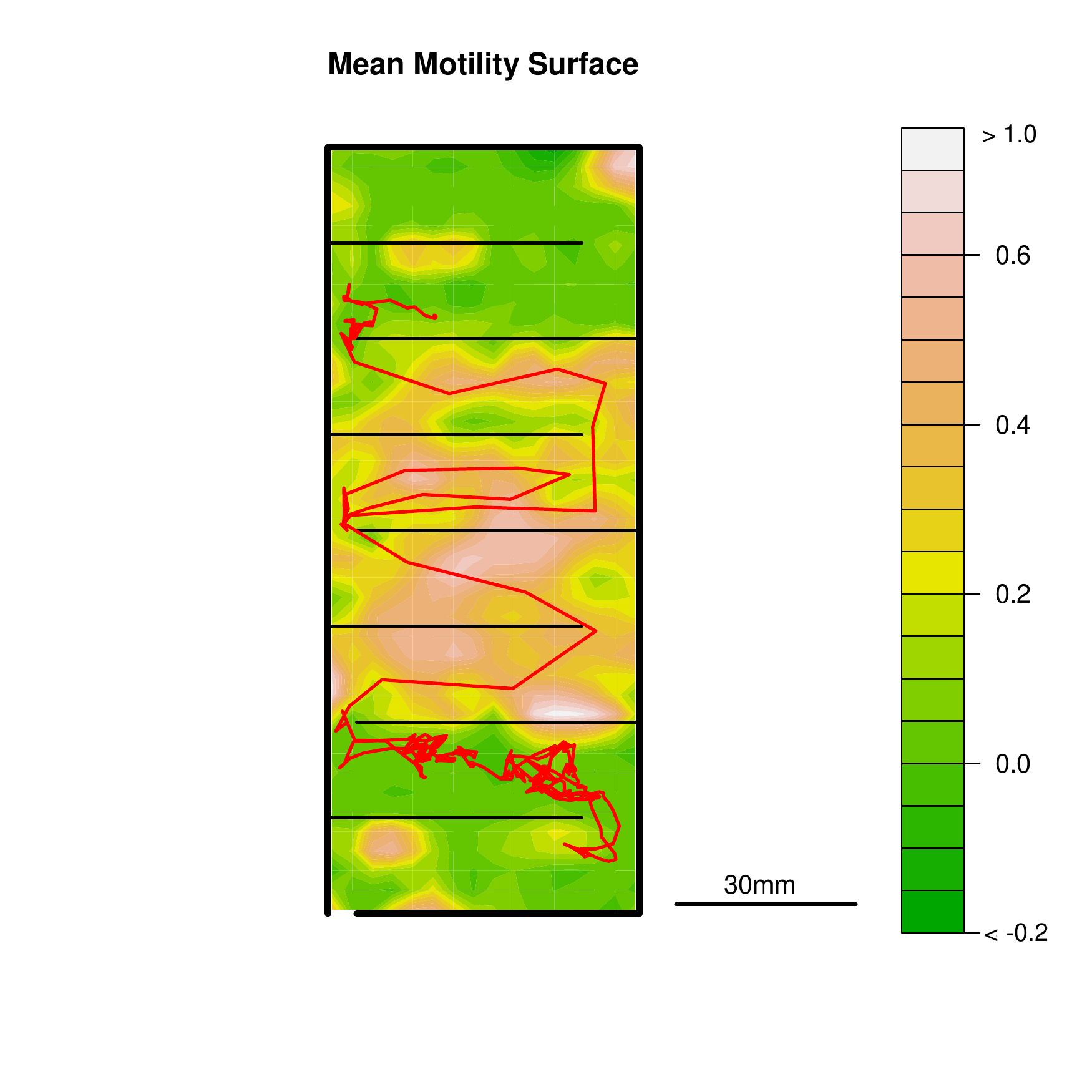}
  \caption{Motility Surface}
\end{subfigure}
\begin{subfigure}{.32\textwidth}
  \centering
  \includegraphics[clip, trim=1.5cm 1.35cm 1.5cm 1.5cm, width=.95\linewidth, page=1]{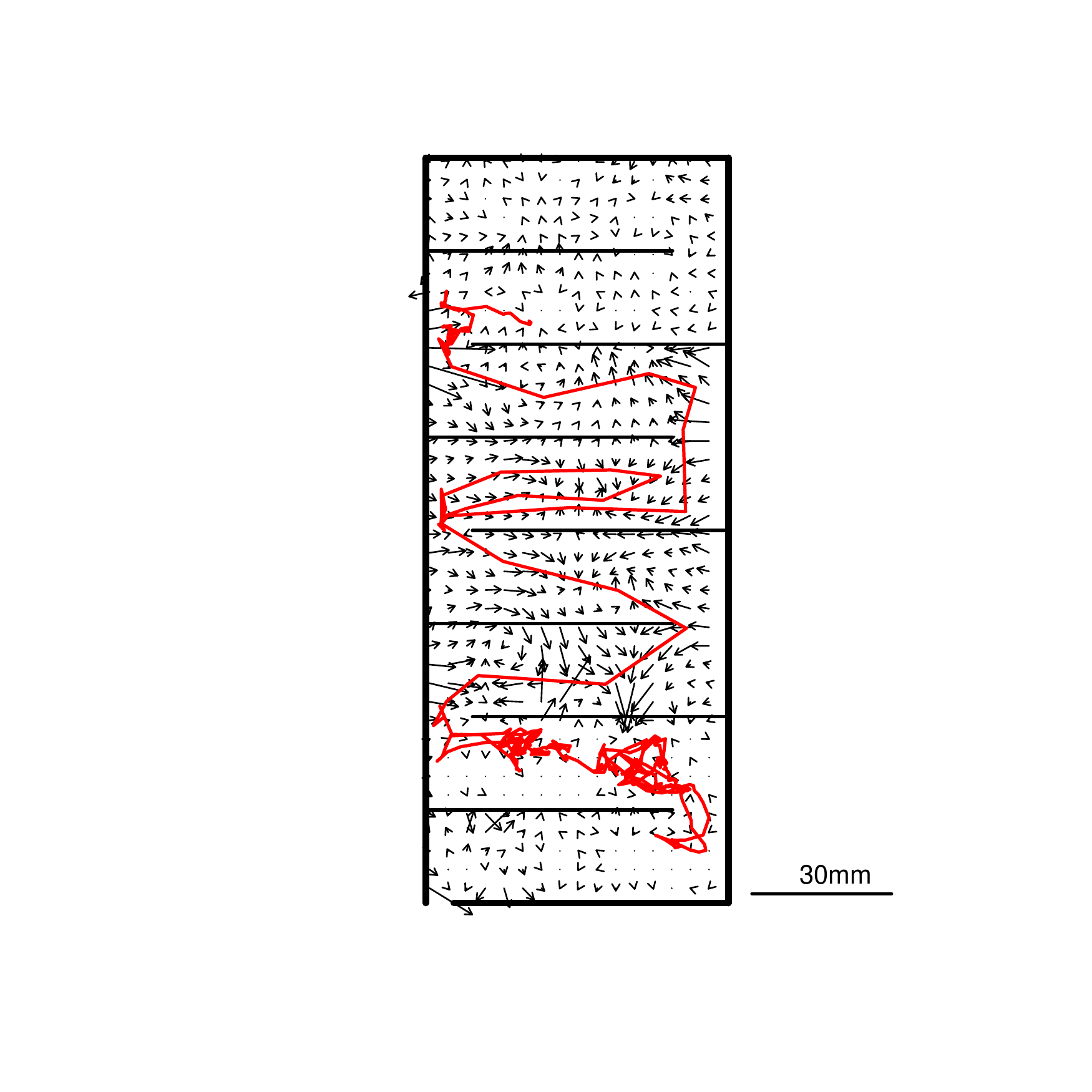}
  \caption{Combined Gradient Field}
\end{subfigure}%
\caption{Ant Posterior Mean Surfaces}
\label{fig:antresults}
\end{figure}

Estimates of the posterior mean potential surface and motility surface are constructed by taking the point-wise posterior mean of each coefficient, and plotting the resulting surfaces over a fine grid. Estimating the variability in the potential surface is complicated by the fact that we are only interested in the relative height. Shifting the level of the entire surface by a constant has no impact on movement since the potential surface only impacts behavior through its gradient. The potential surface, plotted in Figure \ref{fig:antresults} (a), reveals a tendency to move away from the walls on the left and right sides of the nest, particularly in the center chambers (II and III). This is consistent with ants that turn as they approach the walls while traveling between chambers.

The motility surface, plotted in Figure \ref{fig:antresults} (b), reveals a tendency to move faster in the center of the nest, in chambers II and III. This suggests that this area is mainly used by the ants for commuting between chambers. Additionally, there is low relative motility in each of the doorways between chambers. The combined gradient field plotted in Figure \ref{fig:antresults} (c) reveals both the tendency to move away from walls, and the tendency to move faster in the central chambers.

\begin{figure}
\centering
\begin{subfigure}{.5\textwidth}
  \centering
  \includegraphics[clip, trim=1.5cm 1.35cm 0.5cm 0.05cm, width=.95\linewidth, page=2]{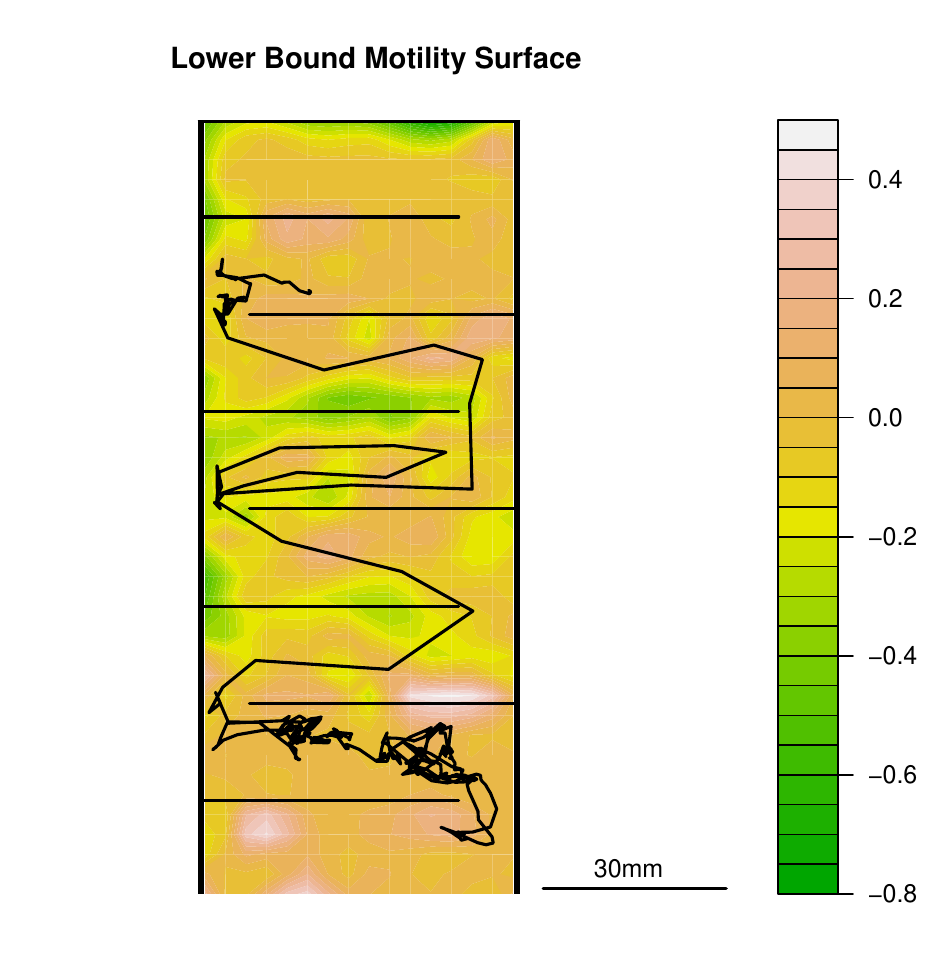}
  \caption{Lower Credible Surface}
\end{subfigure}%
\begin{subfigure}{.5\textwidth}
  \centering
  \includegraphics[clip, trim=1.5cm 1.35cm 0.5cm 0.05cm, width=.95\linewidth, page=3]{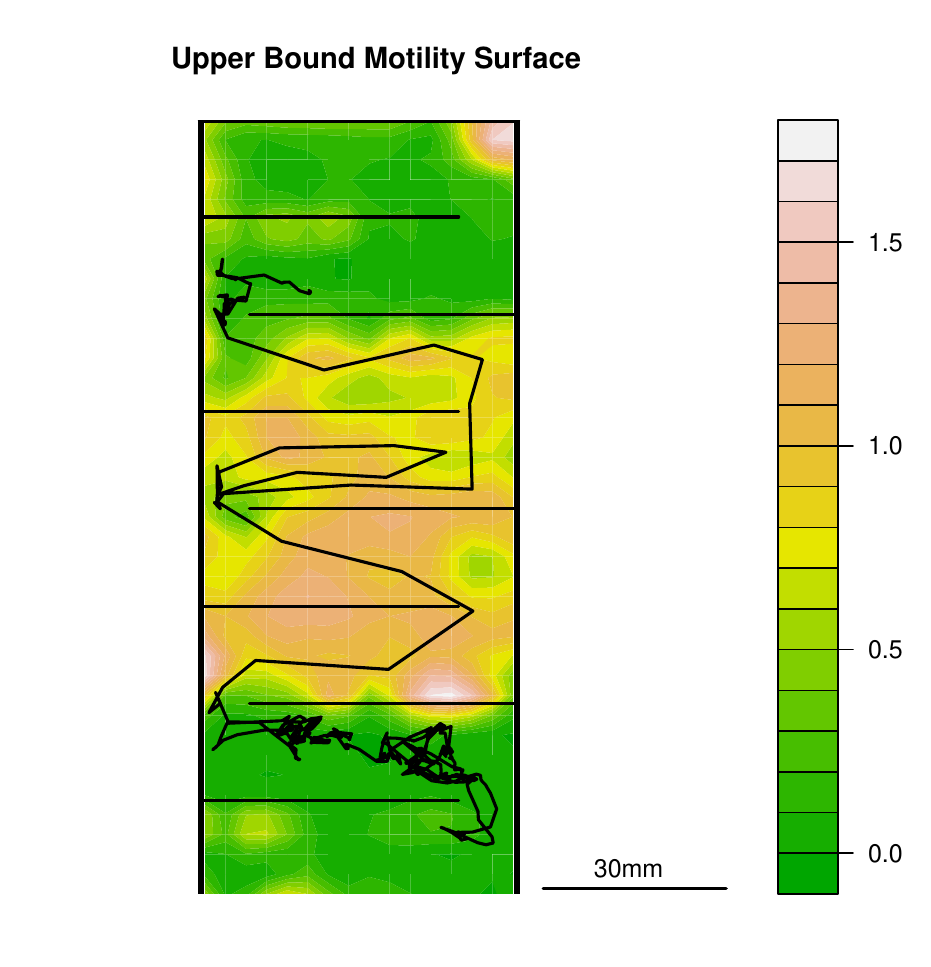}
  \caption{Upper Credible Surface}
\end{subfigure}
\caption{Ant Posterior Motility Surface}
\label{fig:CredibleSurfaces}
\end{figure}

The 95\% point-wise credible intervals for the motility surface estimated from the ant movement observations, along with a black line denoting the movement path of an arbitrarily selected ant, are plotted in Figure \ref{fig:CredibleSurfaces}. The upper limits in most regions in chambers I and IV are below the lower limits in parts of chambers II and III, indicating that the difference in movement rates in the different areas of the nest are statistically significant. Additionally, the regions with low relative motility in the doorways are apparent, particularly in the doorway connecting the chambers I and II and in the doorway connecting chambers II and III. This might be due to the linear interpolation procedure described in Section \ref{data}, or it could reveal a tendency to move slower in the doorways between these chambers. A simpler model for animal movement would not be able to capture this variation, nor would it model the statistically significant autocorrelation presented in Table \ref{table:antParams}.

To assess model fit, the one-step ahead mean prediction error for the ant locations is calculated. At each time step, the next location $\left( \hat{x}_{t_i,j}, \hat{y}_{t_i,j} \right) $ (for all 106,767 observations of ants in the nest) is predicted using the model by drawing predicted locations at each iteration of the Markov chain. The average prediction error, which is the distance between the predicted location and the observed location for each of the 106,767 observations, at each MCMC iteration is calculated using
    \begin{linenomath*}
    \begin{align*}
    \frac{1}{106,767}\sum_i \sum_j \sqrt{(\hat{x}_{t_i,j} - x_{t_i,j})^2 + \hat{y}_{t_i,j} - y_{t_i,j})^2 }
    \end{align*}
    \end{linenomath*}
The result is then averaged over all draws from the Markov chain. The prediction error was estimated for three models, the full model (described above), a model with a constant potential surface ($H(x,y)\equiv 0$), and a model with a constant motility surface ($M(x,y)\equiv 1$). The mean prediction error for each of the models is, respectively 0.09mm, 0.11mm, and 0.64mm. This indicates that the model that estimates both motility and potential surfaces is the best at predicting animal movements, and incorporating the motility surface in the model improves prediction more than the potential surface. This reflects the variability in movement rate in different spatial locations within the nest.

\section{Discussion}\label{discussion}

The SDE based model we have developed allows for autocorrelated movement and flexible spatially-varying drift and velocity. Ant movement observed every second within a nest for an hour is investigated, and the transit behavior of ants between different chambers is captured. The motility surface impacts the rate at which ants spread through the nest. Modelling this is important, as the spatially-varying rate of movement has important implications for the spread of resources, pathogens, and pesticides within the nest. The spatially-varying model presented is used here for experimental data. The applicability of this model to observational data should be considered on a case by case basis.

This model is flexible, allowing for temporal autocorrelation and movement which is dependent on an animal's environment. The environmental dependency can be based on covariates, such as resources in the area, it can be nonparametric, using a variety of basis functions, or it can be a semi-parametric structure combining both environmental features and the additional spatially-varying behavior. For example, our results for the movement of ants indicate that the motility surface plotted in Figure \ref{fig:antresults} (b) could be effectively modeled using dummy covariates for each ``chamber". The ability to estimate the surface using non-parametric methods is important, as it is often impossible to observe all environmental covariates when analyzing animal movement over large areas. Thus this model provides a flexible framework to model movement to help understand a variety of processes. Examples include resource selection, impacts of environmental changes, and the spread of agents, such as pathogens or nutrients through animal societies.

Some parts of the model are only identifiable up to a multiplicative constant. Therefore, interpretation of the estimated surfaces should be limited on the scale of the motility and potential surfaces. Further, computation becomes more burdensome as the number of basis functions and observations increase. Thus, more efficient methods for inference would be necessary to estimate surfaces with  high resolution features, such as walls or fences, in very large regions.
The model requires a high temporal rate of observation of animal locations due to the Euler-Maruyama approximation to the SDEs. To correct this issue, locations between observation times can be imputed from the model, however this may become computationally taxing since the motility and potential surface estimates require using B-spline basis representation for these imputed observations. Updating the basis representation of these imputed locations at every iteration in a Metropolis-Hastings algorithm is often slow. Alternatively, higher order approximations to the SDEs can be used. which may provide more accurate numerical approximations of the underlying SDE model \eqref{eq:SDE1} and \eqref{eq:SDE2}.

In future work, temporally varying behavior will be considered. A state-space model with distinct potential surfaces and movement parameters in different states could be utilized, however these states may be difficult to identify if the behavior in different states is similar \cite{beyer2013effectiveness}. This could also allow for the interaction of ants through combining a latent interaction network model \cite{scharf2015dynamic} with potential surfaces consisting of directed movement toward specific neighboring ants.

\section*{Acknowledgements}
This research is supported by the NSF grant EEID 1414296.

\bibliography{extracted}

\newpage

\appendix
\renewcommand\thefigure{\thesection.\arabic{figure}}

\section{Appendix }
\setcounter{figure}{0}

\subsection{Prior Specification}

We use priors given by:\\
$v_{1,j}^{(x)} \sim N(0, 1000)$\\
$v_{1,j}^{(y)} \sim N(0, 1000)$\\
$\beta \propto \mbox{exp}\left( \frac{-(\beta-1)^2}{20000} \right)I(\beta>0)$\\
$\sigma^2 = 1$\\
$\kappa^2 \sim IG(0.001, 0.001)$\\
$\boldsymbol{\gamma} \sim N \left(\boldsymbol{0}, \left(\tau_{\gamma}\left( \boldsymbol{D} -\rho_{\gamma} \boldsymbol{Q} \right)  \right)^{-1} \right)$\\
$\tau_{\gamma} \sim \mbox{Exp}(\mu_{\alpha}^2)$\\
$\rho_{\gamma} \sim \mbox{Uniform}(0.01,0.99)$\\
$\boldsymbol{\alpha} \sim N \left(\mu_{\alpha} \boldsymbol{1}, \mu_{\alpha}^2 \left(\tau_{\alpha}\left( \boldsymbol{D} -\rho_{\alpha} \boldsymbol{Q} \right) \right)^{-1} \right)$\\
$\mu_{\alpha} \sim \mbox{N}(1,1)$\\
$\tau_{\alpha} = 9$\\
$\rho_{\alpha} \sim \mbox{Uniform}(0.01,0.99)$\\
$r_1 \sim \text{lognormal}(10, 1)$\\
where,\\
$(\boldsymbol{Q})_{kk} = 0 $\\
$(\boldsymbol{Q})_{kl} = I(k,l \mbox{ are neighbors})$\\
$(\boldsymbol{D})_{kk} = \# \mbox{ neighbors of k}$\\
$(\boldsymbol{D})_{kl} = 0 \mbox{ if } k \neq l$

\newpage

\subsection{Full Conditional Distributions}

In this section, full conditional distributions are given for the model Presented in Section \ref{inference}, equation \ref{eq:EM}.

\begin{enumerate}

\item $\beta \sim N_{+}\left( \left(\sum_{j=1}^{N_a} a + \frac{1}{2 \tau_{\beta}}\right) \left(\sum_{j=1}^{N_a} b_j + \mu_{\beta} \right)^{-1}, \left( \sum_{j=1}^{N_a} b_j + \mu_{\beta} \right)^{-1} \right)[0, \infty]$
    \begin{align*}
    a_j=&\frac{\Delta}{2 \sigma^2} \sum_{i=1}^{N_t} \left( -\frac{d}{dx} H(x_{i,j}, y_{i,j}) - v_{i,j}^{(x)} \right)^2 + \\
        &\frac{\Delta}{2 \sigma^2} \sum_{i=1}^{N_t} \left( -\frac{d}{dy} H(x_{i,j}, y_{i,j}) - v_{i,j}^{(y)} \right)^2
    \end{align*}
    \begin{align*}
    b_j=\frac{1}{2 \sigma^2} & \sum_{i=1}^{N_t} \left( -\frac{d}{dx} H(x_{i,j}, y_{i,j}) - v_{i,j}^{(x)} \right) \left( v_{i+1,j}^{(x)}  - v_{i,j}^{(x)}  \right) + \\
        \frac{1}{2 \sigma^2} & \sum_{i=1}^{N_t} \left( -\frac{d}{dy} H(x_{i,j}, y_{i,j}) - v_{i,j}^{(y)} \right) \left( v_{i+1,j}^{(y)}  - v_{i,j}^{(y)}  \right)
    \end{align*}



\item $\kappa^2 \sim IG(\alpha_{\kappa}+ N_t*N_a - 1, \sum_{j=1}^{N_a} a_j + \beta_{\kappa})$

    \begin{align*}
    a_j=\frac{1}{2\Delta} & \sum_{i=1}^{N_t} \left( x_{i+1,j} - x_{i,j} - M(x_{i,j}, y_{i,j}) v_{i,j}^{(x)} \Delta \right)^2 + \\
        \frac{1}{2\Delta} & \sum_{i=1}^{N_t} \left( y_{i+1,j} - y_{i,j} - M(x_{i,j}, y_{i,j}) v_{i,j}^{(y)} \Delta \right)^2
    \end{align*}

\newpage

\item $v^{(x)}_{1,j} \sim N( a_{1,j}^{-1}b_{1,j} , a_{1,j}^{-1} )$ for $j \in \{1,...,N_a \}$

    \begin{align*}
    a_{1,j}=\frac{\Delta}{ \kappa^2} \left( M(x_{1,j}, y_{1,j})\right)^2 + \frac{\beta \Delta -1}{\Delta \sigma^2} + \frac{1}{\tau_{v_{1}}^2}
    \end{align*}
    \begin{align*}
    b_{1,j}=\frac{1}{\kappa^2} M(x_{1,j}, y_{1,j}) \left(x_{2,j} - x_{1,j}  \right) - \frac{\beta \Delta - 1}{ \Delta  \sigma^2}\left( v_{2,j}^{(x)} + \beta \Delta \frac{d}{dx} H(x_{1,j}, y_{1,j})\right)
    \end{align*}

\item $v^{(x)}_{i,j} \sim N( a_{i,j}^{-1}(b_{i,j}) , a_{i,j}^{-1} )$ for $i \in \{2,...,N_t-1 \}, j \in \{1,...,N_a \}$

    \begin{align*}
    a_{i,j}=\frac{\Delta}{ \kappa^2} \left( M(x_{i,j}, y_{i,j})\right)^2 + \frac{\beta \Delta -1}{\Delta \sigma^2} + \frac{1}{\Delta \sigma^2}
    \end{align*}
    \begin{align*}
    b_{i,j}=\frac{1}{\kappa^2} M(x_{i,j}, y_{i,j}) \left(x_{i+1,j} - x_{i,j}  \right) - \frac{\beta \Delta - 1}{ \Delta  \sigma^2}\left( v_{i+1,j}^{(x)} + \beta \Delta \frac{d}{dx} H(x_{i,j}, y_{i,j})\right) \\
     + \frac{1}{\Delta \sigma^2}\left( v_{i-1,j}^{(x)} + \beta \Delta \left( \frac{d}{dx} H(x_{i-1,j}, y_{i-1,j}) - v_{i-1,j}^{(x)}\right) \right)
    \end{align*}

\item $v^{(x)}_{N_t,j} \sim N( a^{-1}b_{N_t,j} , a^{-1} )$ for $j \in \{1,...,N_a \}$

    \begin{align*}
     a=\frac{1}{\Delta \sigma^2}
    \end{align*}
    \begin{align*}
    b_{N_t,j}=\frac{1}{\Delta \sigma^2}\left( v_{N_t-1,j}^{(x)} + \beta \Delta \left( \frac{d}{dx} H(x_{N_t-1,j}, y_{N_t-1,j}) - v_{N_t-1,j}^{(x)}\right) \right)
    \end{align*}

\newpage

\item $\boldsymbol{\gamma} \sim N(\cdot, \cdot)$ with mean $\left( \sum_{j=1}^{N_a} \boldsymbol{a}_{i,j} + \tau_{\gamma} \left( \boldsymbol{D} -\rho_{\gamma} \boldsymbol{Q} \right) \right)^{-1} \left(\sum_{j=1}^{N_a} \boldsymbol{b}_{i,j} \right)$ and variance \\ $\left( \sum_{j=1}^{N_a} \boldsymbol{a}_{i,j} + \tau_{\gamma} \left( \boldsymbol{D} -\rho_{\gamma} \boldsymbol{Q} \right) \right)^{-1}$

    \begin{align*}
     \boldsymbol{a}_{i,j}=\frac{\beta^2 \Delta}{\sigma^2} &\sum_{i=1}^{N_t-1} \Phi'_x (x_{i,j}, y_{i,j})\Phi_x (x_{i,j}, y_{i,j}) + \\
     \frac{\beta^2 \Delta}{\sigma^2} &\sum_{i=1}^{N_t-1} \Phi'_y (x_{i,j}, y_{i,j})\Phi_y (x_{i,j}, y_{i,j})
    \end{align*}
    \begin{align*}
    \boldsymbol{b}_{i,j}=\frac{\beta}{\sigma^2} &\sum_{i=1}^{N_t-1} \Phi'_x (x_{i,j}, y_{i,j}) \left( v_{i+1,j}^{(x)} - v_{i,j}^{(x)} + \beta \Delta \left( v_{i,j}^{(x)} +\frac{d}{dx}R(x_{i,j}, y_{i,j})  \right) \right) +\\
    \frac{\beta}{\sigma^2} &\sum_{i=1}^{N_t-1} \Phi'_y (x_{i,j}, y_{i,j}) \left( v_{i+1,j}^{(y)} - v_{i,j}^{(y)} + \beta \Delta \left( v_{i,j}^{(y)} +\frac{d}{dy}R(x_{i,j}, y_{i,j})  \right) \right)
    \end{align*}

    Where $\Phi_x (x_{i,j}, y_{i,j})$ and $\Phi_y (x_{i,j}, y_{i,j})$ are the derivatives of the B-spline Basis functions $\Phi_ (x_{i,j}, y_{i,j})$ with respect to x and y.

\item $\boldsymbol{\alpha} \sim N(\cdot, \cdot)$ with mean $\left(\sum_{j=1}^{N_a} \boldsymbol{a}_{i,j} + \frac{\tau_{\alpha}}{\mu_{\alpha}^2} \left( \boldsymbol{D} -\rho_{\alpha} \boldsymbol{Q} \right) \right)^{-1} \left(\sum_{j=1}^{N_a} \boldsymbol{b}_{i,j} + \frac{\tau_{\alpha}}{\mu_{\alpha}} \left( \boldsymbol{D} -\rho_{\alpha} \boldsymbol{Q} \right)\right)$ and variance $\left(\sum_{j=1}^{N_a} \boldsymbol{a}_{i,j}+ \frac{\tau_{\alpha}}{\mu_{\alpha}^2} \left( \boldsymbol{D} -\rho_{\alpha} \boldsymbol{Q} \right)\right)^{-1}$

    \begin{align*}
     \boldsymbol{a}_{i,j} =\frac{\Delta}{\kappa^2} &\sum_{i=1}^{N_t-1} (v_{i,j}^{(x)})^2 \Phi'_x (x_{i,j}, y_{i,j})\Phi_x (x_{i,j}, y_{i,j}) + \\
     \frac{\Delta}{\kappa^2} &\sum_{i=1}^{N_y-1} (v_{i,j}^{(y)})^2 \Phi'_y (x_{i,j}, y_{i,j})\Phi_y (x_{i,j}, y_{i,j})
    \end{align*}
    \begin{align*}
    \boldsymbol{b}_{i,j} =\frac{1}{\kappa^2} &\sum_{i=1}^{N_t-1} \Phi'_x (x_{i,j}, y_{i,j}) v_{i,j}^{(x)} \left( x_{i+1,j} - x_{i,j} \right) + \\
    \frac{1}{\kappa^2} &\sum_{i=1}^{N_t-1} \Phi'_y (x_{i,j}, y_{i,j}) v_{i,j}^{(y)} \left( y_{i+1,j} - y_{i,j} \right)
    \end{align*}

\item $\tau^2_{\gamma} \sim IG( \frac{1}{2}N_h * N_M + \alpha_{\tau_{\gamma}}, \frac{1}{2}\boldsymbol{\gamma}' \left( \boldsymbol{D} -\rho_{\gamma} \boldsymbol{Q} \right) \boldsymbol{\gamma} + \beta_{\tau_{\gamma}} )$

\end{enumerate}

\newpage


\subsection{Analysis of the Euler-Maruyama Approximation Error}

\begin{figure}
\centering
\begin{subfigure}{.5\textwidth}
  \centering
  \includegraphics[width=.9\linewidth, page=4]{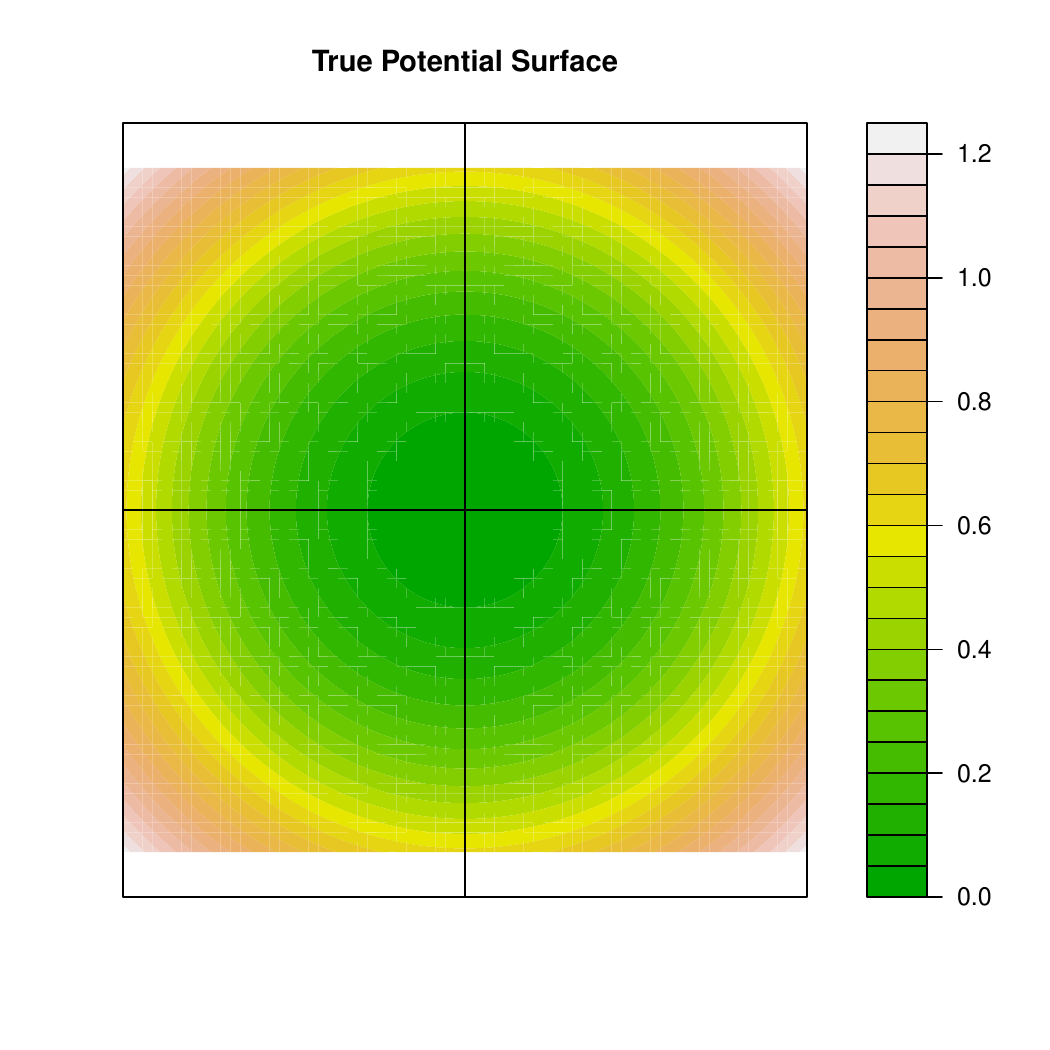}
\end{subfigure}%
\begin{subfigure}{.5\textwidth}
  \centering
  \includegraphics[width=.9\linewidth, page=4]{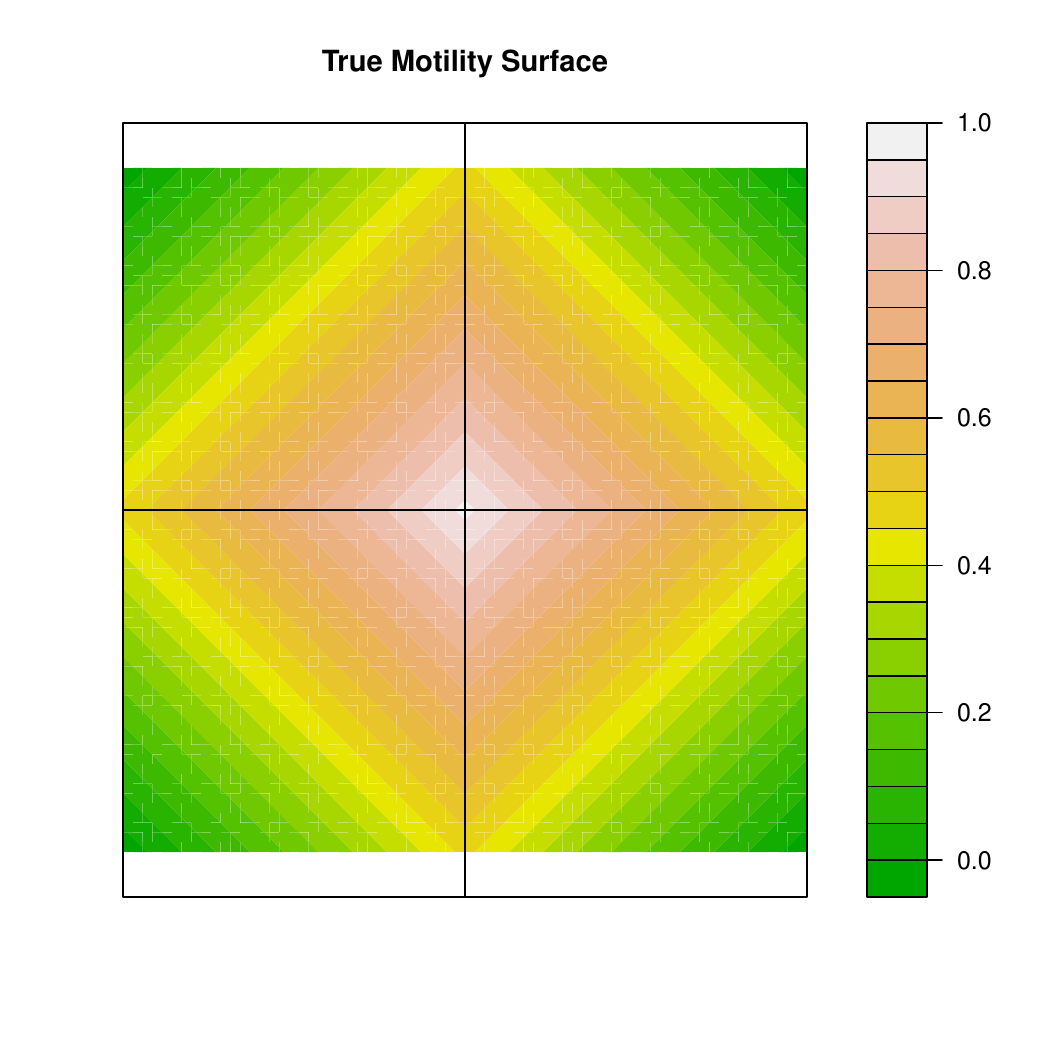}
\end{subfigure}
\caption{True Surfaces for the Thinned Simulated Data Examples.}
\label{fig:Thin_truSurf}
\end{figure}

The Euler-Maruyama method discussed in Section \ref{inference} only provides an accurate approximation of the continuous time process when the temporal increments of the discrete time approximation are sufficiently small. Determination of an acceptable temporal resolution depends on the underlying problem. Various simulations should be used to gain an understanding of the impact of the discrete approximation \cite{kloeden1992numerical}. In this Section, to gain a better understanding of the impact of the discrete time approximation, we apply our algorithm for Bayesian inference to simulated data sets under a variety of temporal resolutions. Data are simulated from a model with parameters $(\beta=0.8, \sigma^2=1 ,\kappa^2=0.0001)$ and surfaces $H(x,y)=0.05x^2+0.05y^2$,  $M(x,y)=[1-0.25(|x|+|y|)]_+$ with $\Delta=0.1$ for $10$ individuals. The surfaces are plotted in Figure \ref{fig:Thin_truSurf}. Simulated data ``observations" for each individual are dropped so that the impact of the temporal resolution of observation can be studied. For example, if every other observation is dropped, this is equivalent to a discrete time approximation with $\Delta=0.2$. The length of the total observation time is adjusted so that under each scenario, the total number of observations remains the same ($N_t=5000$).

Results for the relevant model parameters are presented in Table \ref{table:SVSDE_APPROX}. The temporal dependence in the model is described by the product $\beta \Delta$. A value of $\beta \Delta=1$ indicates independent movement and the autocorrelation increases as $\beta \Delta \rightarrow 0$. The autocorrelation parameter estimate reveals that we are able to capture the true value of $\beta$ when there is no temporal thinning. As the simulated data is thinned the parameter estimates indicate that as the temporal resolution of observation becomes coarser the dependence between observations is biased towards independence.  If the time lag between observations is large enough, the value of $\beta \Delta$ exceeds $1$, indicating negative correlation in velocity through time. This is closely related to the incorrect scaling of the surfaces, and is discussed further after the surface results are presented. The estimate for variability in location $\kappa^2$ also increases as the temporal resolution increases.

\begin{table}
\begin{center}
  \caption{Computational Efficiency and Surface Resolution}\label{table:SVSDE_APPROX}
\begin{tabular}{ |l |c |c | c | c| }
    \hline
    $\Delta$ & $\beta \Delta$ & 95 \% CI for $\beta \Delta$ & $\kappa^2 \Delta$ & 95 \% CI for $\kappa^2 \Delta$\\ \hline
    $0.1$ & $0.080$ & $(0.077,0.084)$  & $0.00018$ & $(0.00017,0.00019)$ \\ \hline
    $0.2$ & $0.125$ & $(0.121,0.129)$   & $0.00027$ & $(0.00024,0.00030)$ \\ \hline
    $0.5$ & $0.216$ & $(0.210,0.222)$   & $0.0028$ & $(0.0027,0.0029)$ \\ \hline
    $1$ & $0.490$ & $(0.483,0.496)$   & $0.0047$ & $(0.0042,0.0051)$ \\ \hline
    $3$ & $1.122$ & $(1.110,1.134)$   & $0.034$ & $(0.032,0.036)$ \\ \hline
    $5$ & $1.503$ & $(1.485,1.530)$   & $0.051$ & $(0.047,0.055)$ \\ \hline
    $50$ & $2.329$ & $(2.297,2.353)$   & $0.437$ & $(0.401,0.459)$ \\ \hline
\end{tabular}
\end{center}
\end{table}

\begin{figure}
\centering
\begin{subfigure}{.5\textwidth}
  \centering
  \includegraphics[width=.9\linewidth, page=1]{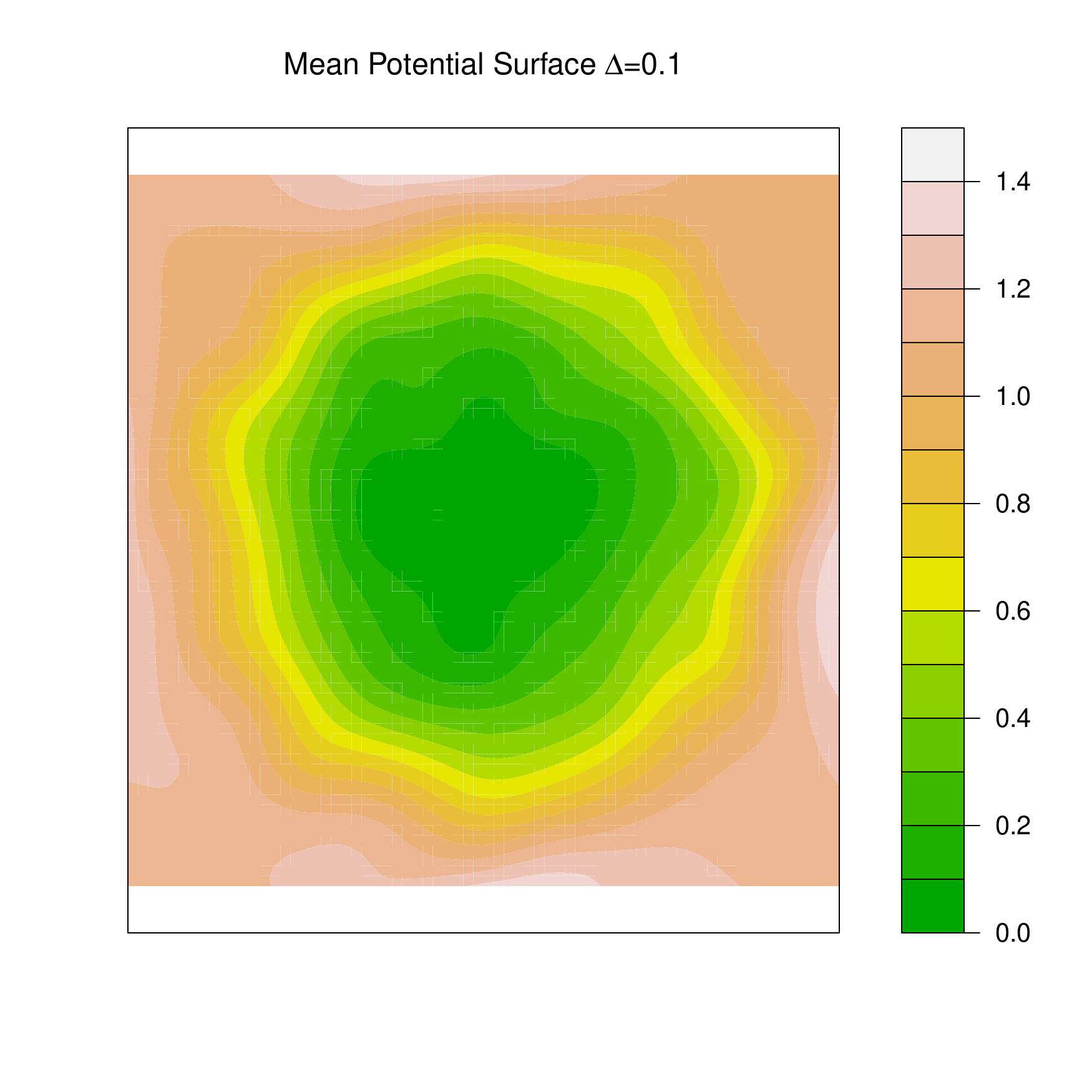}
\end{subfigure}%
\begin{subfigure}{.5\textwidth}
  \centering
  \includegraphics[width=.9\linewidth, page=1]{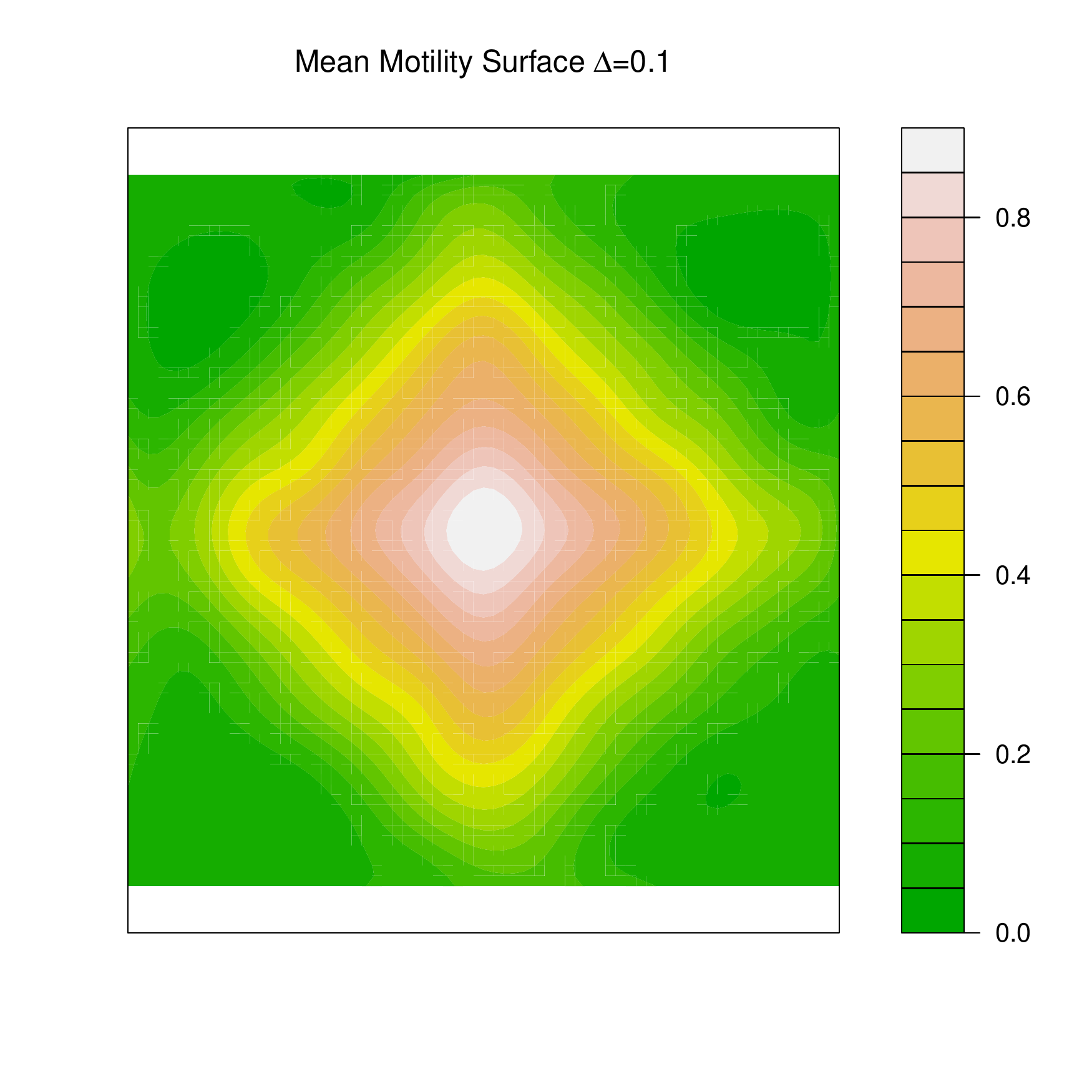}
\end{subfigure}
\begin{subfigure}{.5\textwidth}
  \centering
  \includegraphics[width=.9\linewidth, page=3]{Thin_Potential_Results}
\end{subfigure}%
\begin{subfigure}{.5\textwidth}
  \centering
  \includegraphics[width=.9\linewidth, page=3]{Thin_Motility_Results}
\end{subfigure}
\begin{subfigure}{.5\textwidth}
  \centering
  \includegraphics[width=.9\linewidth, page=1]{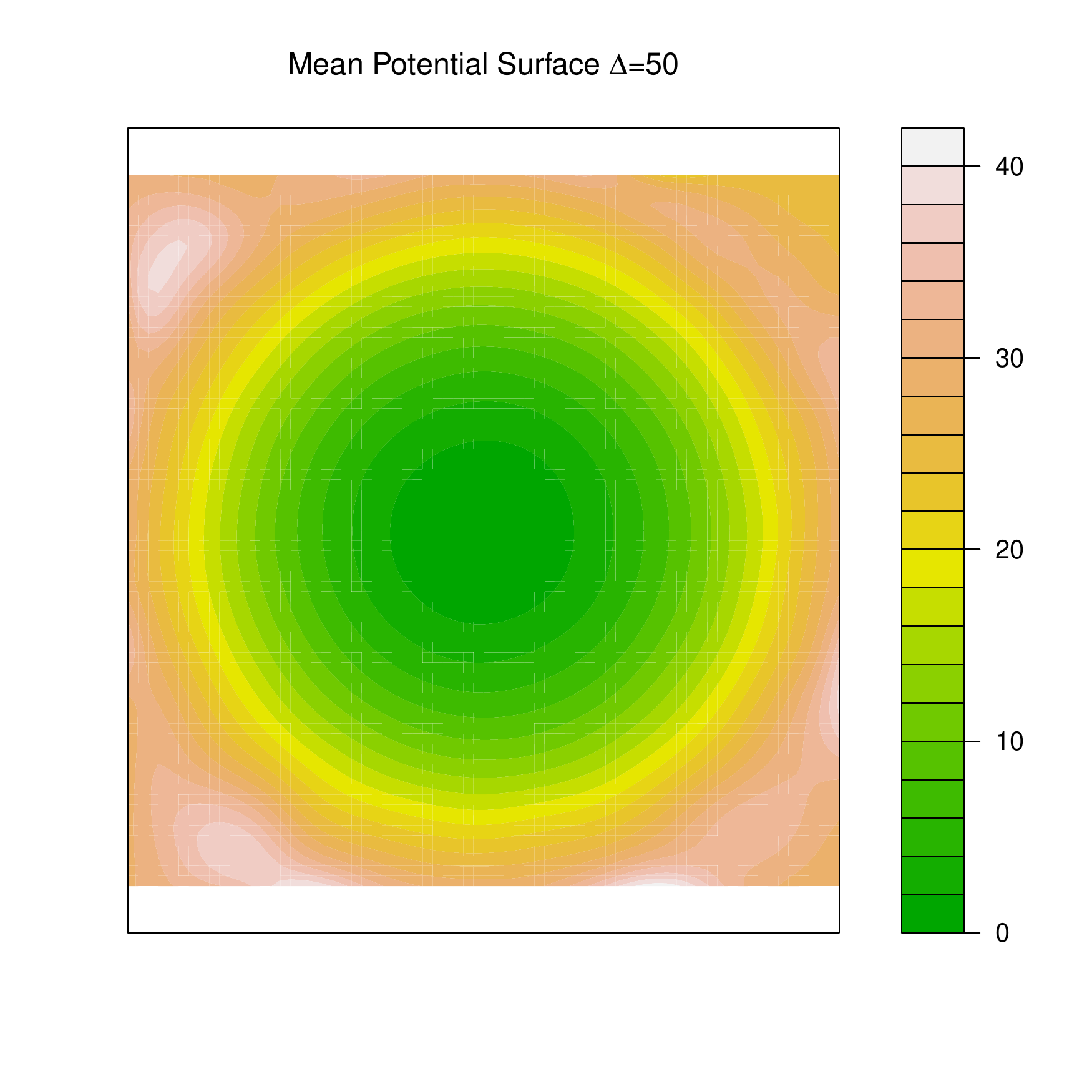}
\end{subfigure}%
\begin{subfigure}{.5\textwidth}
  \centering
  \includegraphics[width=.9\linewidth, page=1]{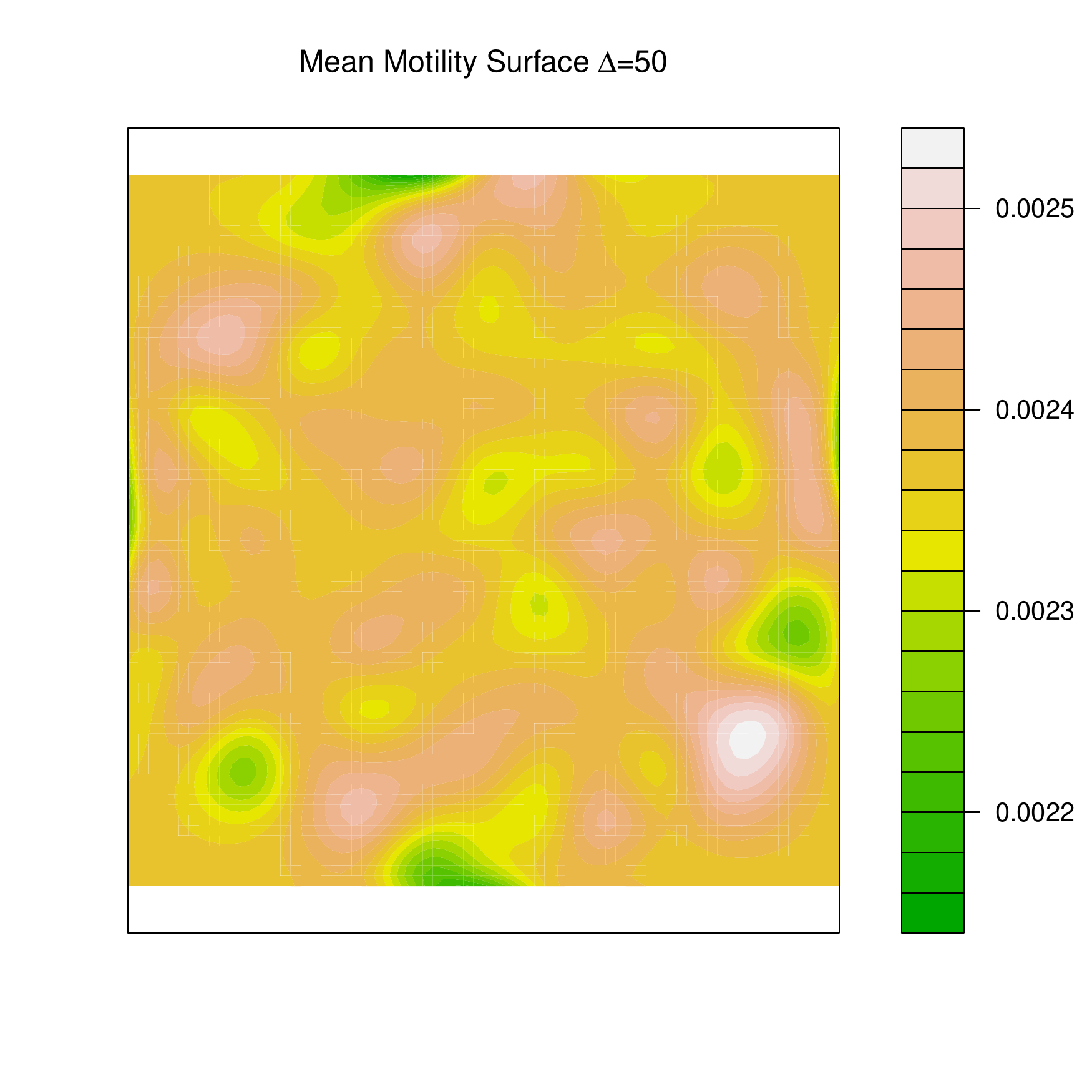}
\end{subfigure}
\caption{Estimated Surfaces for the Thinned Simulated Data Examples.}
\label{fig:Thin_estSurf}
\end{figure}

Estimates for the gradient fields are calculated by taking the point-wise mean of each basis function coefficient as the estimated value and plotting the surface over a fine grid. Results for the potential and motility surfaces in several cases, $\Delta=0.1$ (keeping every simulated observation), $\Delta=5$ (keeping every $50$th simulated observation) and $\Delta=50$ (keeping every $500$th simulated observation) are plotted in Figure \ref{fig:Thin_estSurf}. The results indicate that the change in temporal resolution causes difficulties in estimating the scale of the surfaces. This is related to the incorrect estimate for $\beta \Delta$ described above. Due to identifiability, as discussed in Section \ref{identify}, the value of $\sigma^2$ has been fixed at one. The velocity equation approximation (in one dimension) is therefore
\begin{equation*}
\begin{aligned}
 v^{x}_{t_{i+1}} &= v^{x}_{t_{i}} + \beta \left( - \nabla H\left(x_{t_{i}},y_{t_{i}}\right)  - v^{x}_{t_{i}}\right) \Delta + \epsilon_{x}
\end{aligned}
\end{equation*}
where $ \epsilon_{x}$ is an independent Gaussian random variable with mean $0$ and variance $\Delta$. Thus, the variability in velocity increases with larger values of $\Delta$. This is intuitive for small values of $\Delta$, as the variability in velocity is likely to increase as the time between observations increases. However, as the temporal resolution becomes more coarse, the velocity approximations lose their meaning. For example, if $\Delta=5$, the variability in velocity indicates that speeds greater than $5$ units per observation are reasonable. However, the space use of any individual is limited to the area near the origin by the quadratic well potential surface, such that the maximum distance in the $x$-dimension between any two points in a simulated individual's observed path is less than $5$. Hence, at a fine temporal resolution the latent velocities increase to unrealistic values.

The unrealistic estimates of the latent velocities ($v^{x}_{t}$) result in small estimates for the motility surface (the scale of the surface is underestimated to reduce the movement rate to a realistic value) larger estimates in the gradient for the potential surface (the scale of the surface is overestimated to induce directional bias in the inflated velocities) and incorrect estimates for $\beta*\Delta$ (which determines the weights between the potential surface and the previous velocity in the distribution for the current velocity) driven by the incorrect scaling of the velocities and the potential surface.

The inability to recover the scale of the surfaces is reflected in Figure \ref{fig:Thin_truSurf}. However, the shape of both estimated surfaces remains accurate when $\Delta=5$, as we are able to capture the tendency of individuals to be drawn to the origin via the potential well, and to move faster while near the origin via the motility surface in the simulated data example with the coarsest temporal resolution. If the scale of observation is increased further ($\Delta=50$) the estimated potential surface still retains the shape of a quadratic well but the features of the motility surface disappear. This is intuitive since estimates of differences in movement rate are unreliable when a simulated particle's velocity is averaged over longer periods of time, however directional bias is still reflected in the space-use of an individual at arbitrary time scales.

Overall, Table \ref{table:SVSDE_APPROX} indicates that the estimates for model parameters gradually break down as the temporal resolution increases, however Figure \ref{fig:Thin_estSurf} indicates we are still able to estimate the shape of the potential surface that determines spatial-variation of directional bias in movement behavior when only infrequent observations are available.

\subsection{Application to Simulated Data}

\begin{figure}
\centering
\begin{subfigure}{.45\textwidth}
\flushleft
  \caption{}
  \centering
  \includegraphics[width=.7\linewidth]{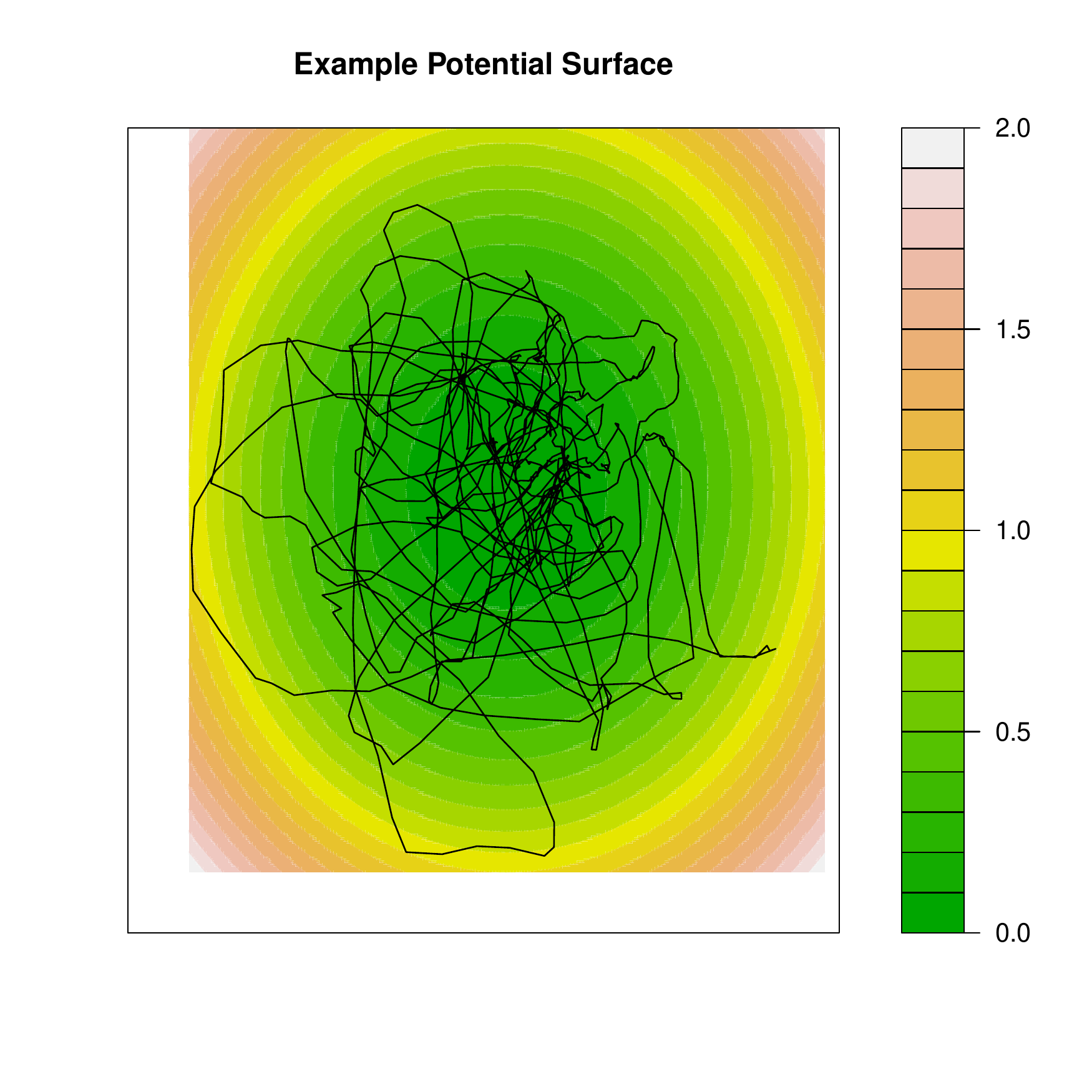}
\end{subfigure}
\begin{subfigure}{.45\textwidth}
  \caption{}
  \centering
  \includegraphics[width=.7\linewidth]{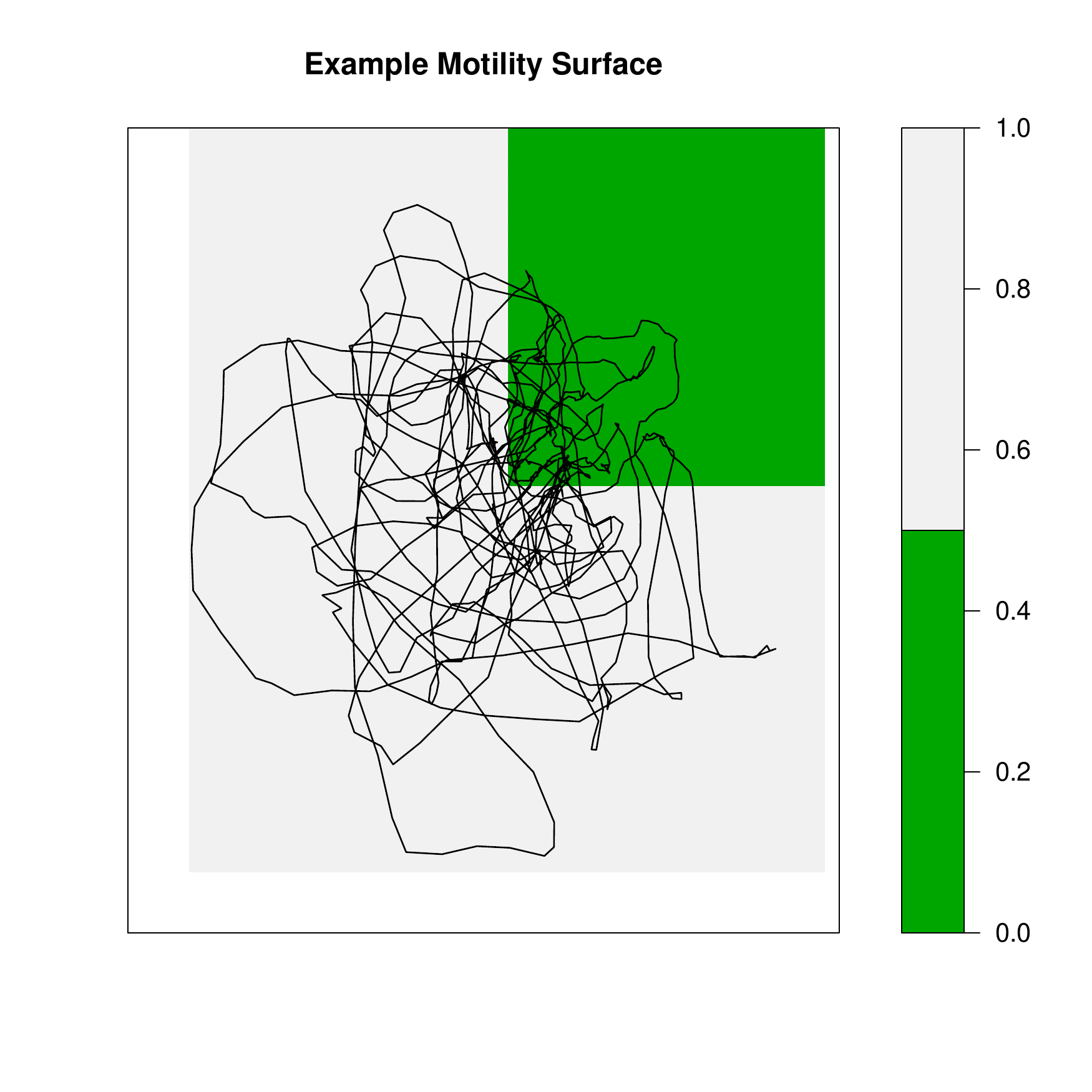}
\end{subfigure}
\begin{subfigure}{.45\textwidth}
  \caption{}
  \centering
  \includegraphics[width=.7\linewidth]{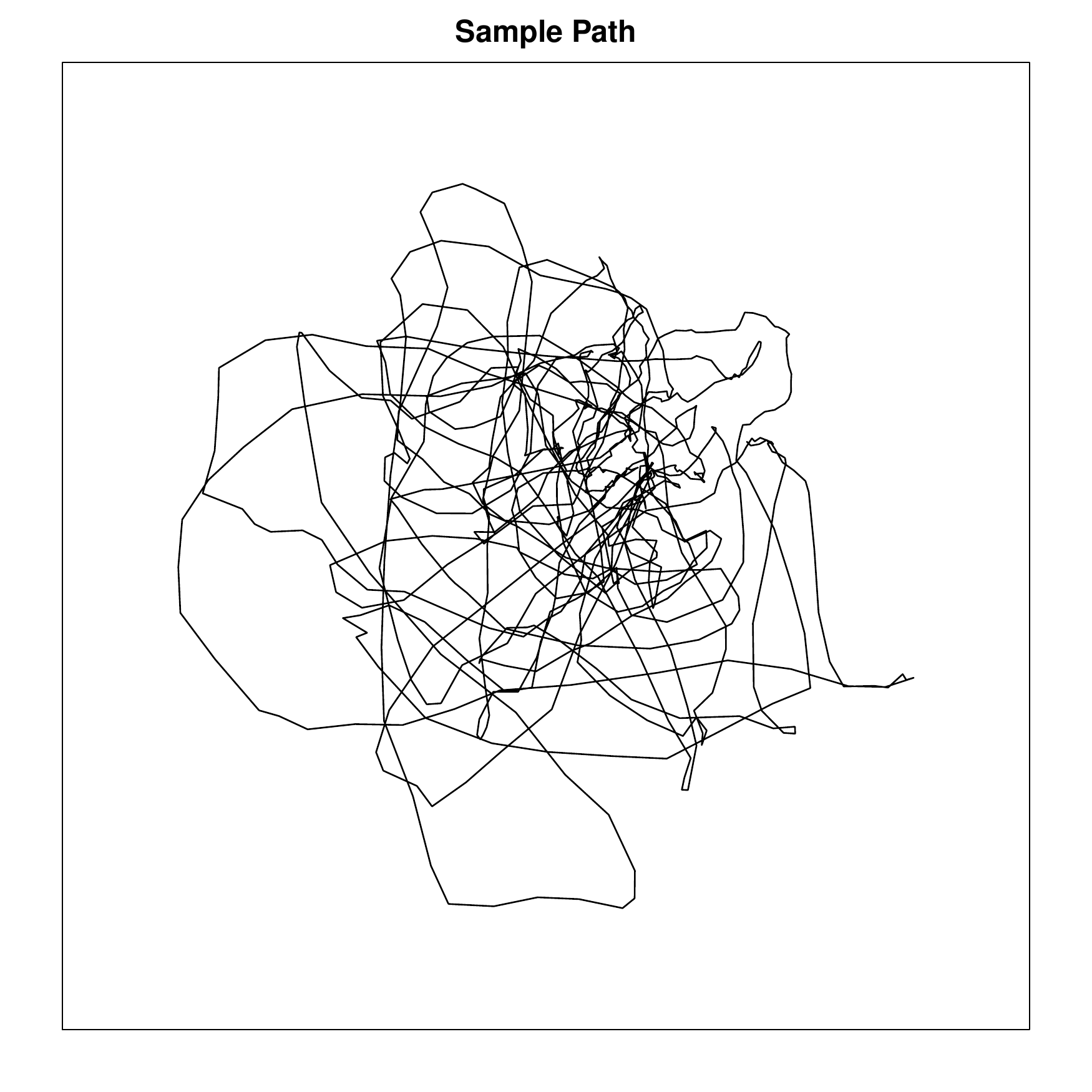}
\end{subfigure}
\begin{subfigure}{.45\textwidth}
  \caption{}
  \centering
  \includegraphics[width=.7\linewidth]{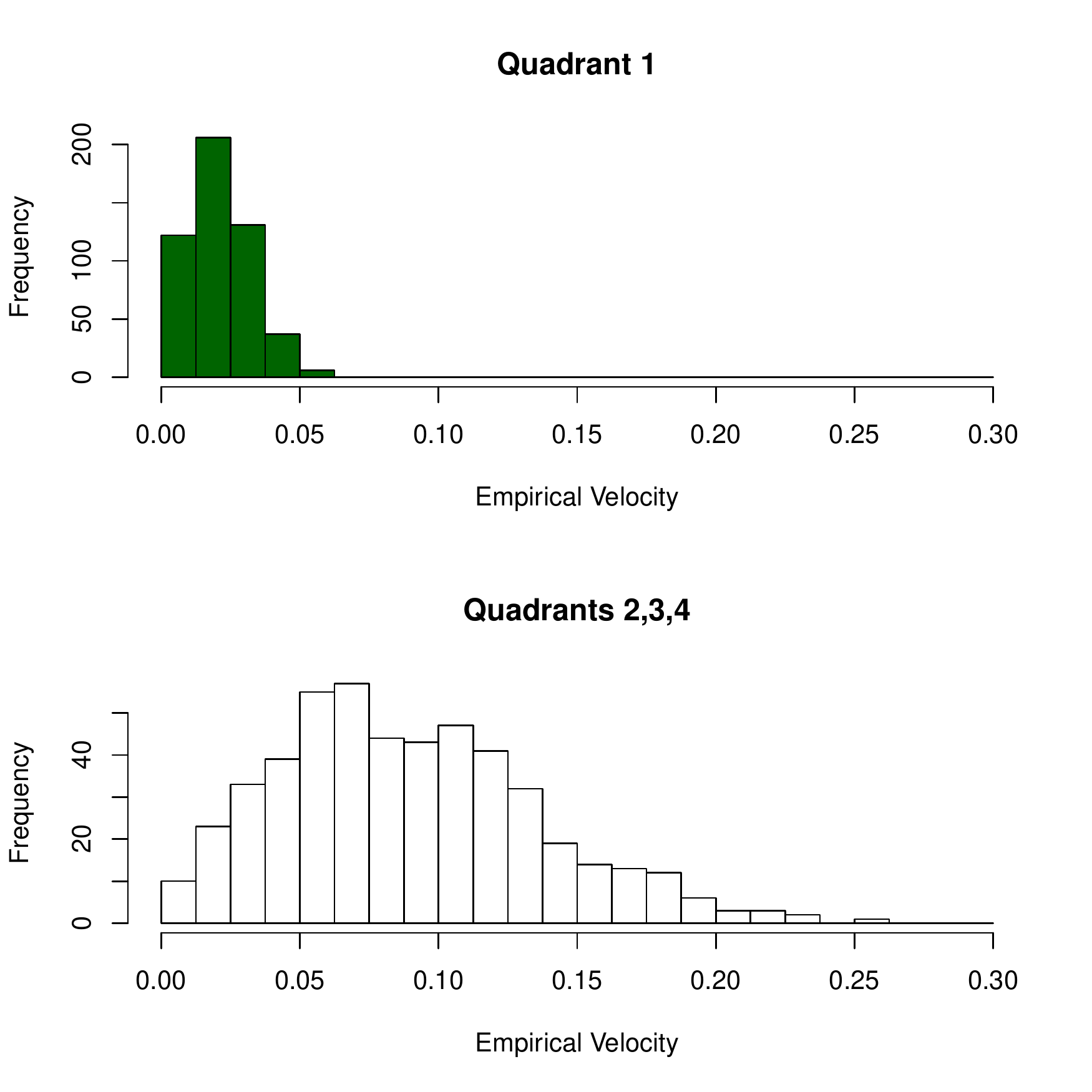}
\end{subfigure}
\caption{(a) The potential function used for the simulation example. (b) The motility function used for the simulation example. (c) A portion of the simulated path for one individual. (d) Histograms of the empirical velocities of the simulated individual while located in quadrant I compared to the other quadrants.}
\label{fig:test}
\end{figure}

To illustrate the range of behavior possible under our SDE model, we generated a 2-dimensional simulation using the Euler-Maruyama method described in Section \ref{inference}. For simplicity, a quadratic well with a center at the origin is used for the potential surface ($H(x,y)=x^2+y^2$). For the motility surface, the upper right quadrant is set to $0.25$ and the remainder of the support is set to $1$.
 \begin{equation}
 M(x,y)=\begin{cases}
      0.25 & x > 0, y >0  \\
      1 & x<0 \mbox{ or } y<0
   \end{cases}
 \end{equation}
 The potential surface and the motility surface are plotted in Figure \ref{fig:test} (a) and (b) respectively.

 The model parameters were set to the following values: ($\beta=1.5 $ , $\sigma^2=1$ , $\kappa^2=0.01$, $m_1=0$). A total of 6000 observations were simulated for five individuals, with observations made every 0.1 units of time. The first 2000 locations of the simulated data for one of the individuals is plotted in Figure \ref{fig:test}(c). In the simulation, the animal tends to move around a central point of attraction at the origin. The path is somewhat smooth due to the directional persistence included in the model. The histograms plotted in Figure \ref{fig:test}(d) indicate the animal moves at a slower speed in the first quadrant, consistent with the specification of the motility surface. This simulated movement path demonstrates that our model allows for spatially-varying movement rate as well as directional persistence and spatially-varying drift. This is necessary for the ants, whose movement rate depends on spatial location, as illustrated in Figure \ref{fig:EDA} (b). Previous models allow for directional persistence or spatially-varying drift, whereas our model is flexible enough to combine these directional traits with spatially-varying absolute movement rate and allow for different movement behavior in different chambers of the nest.

To gain a better understanding of our model, inference is performed on the simulated data example. The posterior sample means and $95$\% credible intervals for several parameters are presented in Table \ref{table:sim1Params}.
\begin{table}
\begin{center}
  \caption{Simulated Data}\label{table:sim1Params}
  \begin{tabular}{| l | c  | c | c | c | c | c |}
    \hline
    Parameter  & truth  & posterior mean & credible interval \\ \hline
    $\beta$    & $1.5$  & $1.496$         & $(1.406, 1.587) $    \\
    $\kappa^2$ & $0.01$ & $0.0109$          & $(0.0106, 0.0112)$  \\
    \hline
  \end{tabular}\\
\end{center}
\end{table}
The resulting posterior means and credible intervals indicate that we are able to recover the true values (when true values exist).

\begin{figure}
\centering
\begin{subfigure}{.5\textwidth}
  \centering
  \includegraphics[width=.9\linewidth]{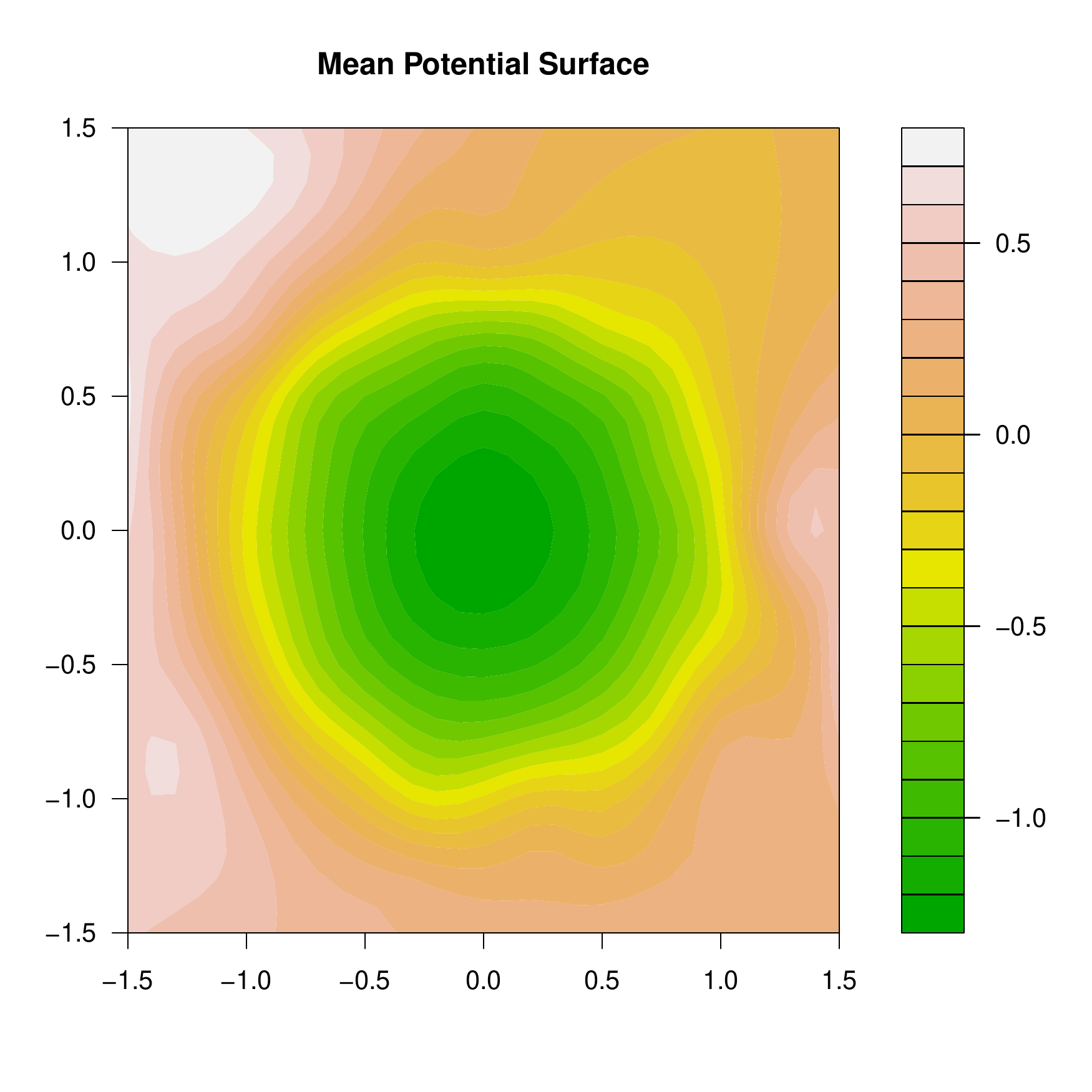}
\end{subfigure}%
\begin{subfigure}{.5\textwidth}
  \centering
  \includegraphics[width=.9\linewidth]{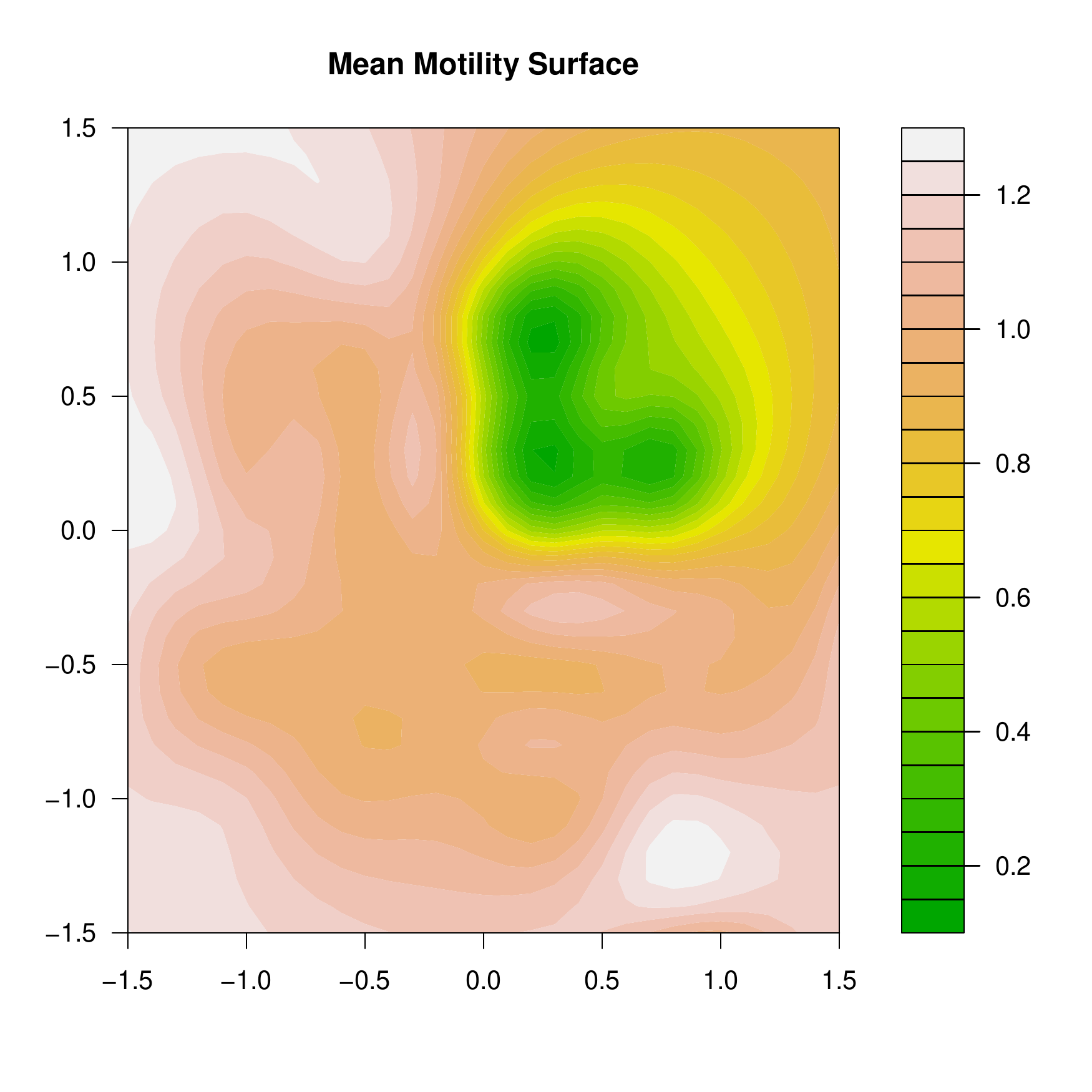}
\end{subfigure}
\caption{Posterior mean surface estimates for the simulated data example.}
\label{fig:simest}
\end{figure}

Estimates of the potential surface and motility surface are plotted in Figure \ref{fig:simest} (a) and (b) respectively. These surfaces are constructed by taking the mean of each coefficient in the joint posterior distribution.
 As evident in Figure \ref{fig:simest} (a), the estimated posterior surface captures the tendency to draw particles back to the origin. The height of the potential surface is arbitrary, since the potential surface only impacts movement through it's gradient, and therefore the height is inestimable. The resulting posterior mean motility surface plotted in Figure \ref{fig:simest} (b) matches the truth well, as it captures the shape and height of the surface accurately in areas where simulated data is observed.

 \newpage

 \subsection{Application to Spread of pathogens}

\begin{figure}
\centering
\begin{subfigure}{.99\textwidth}
  \centering
  \includegraphics[ width=0.16\textwidth]{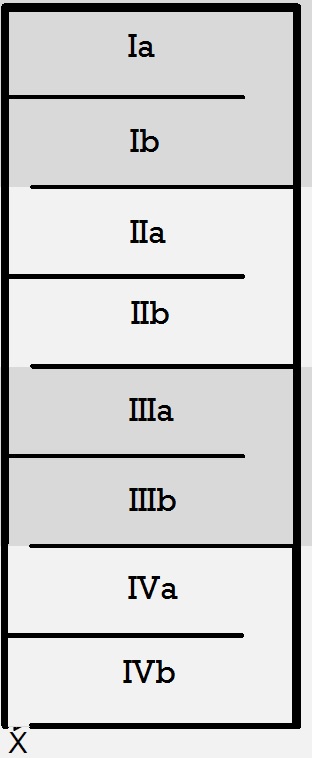}
\end{subfigure}%
\caption{Simulation Nest Layout}
\label{fig:SimNestSpread}
\end{figure}

 The spread of nutrients, pathogens, and other agents is important to the health of a community of ants, as well as other group living organisms, including humans. The spread of such agents is closely related to the movement of ants in the nest. To study the impact of the spatially-varying movement behavior of ants, we simulated the spread of 100 ants, each starting at the exit doorway located at the bottom left corner of chamber IV (Figure \ref{fig:SimNestSpread}). The parameters for the forward simulation were set to the mean values from the posterior distributions. In the movement corridors, the movement of ants between one second intervals is large enough so that they can pass through walls without encountering the wall repulsion effect. A finer temporal resolution would be necessary to capture and recreate movement behavior that doesn't exhibit movement through the walls. In the simulation, movements were truncated to prevent ants from traveling through walls. To estimate the time it would take for an agent to spread to different regions of the nest, the nest was split into 8 sections, two in each chamber. This was done by dividing each of the 4 chambers into two halves along the wall across the middle of the nest. The resulting sections have been designated Ia, Ib, ... , IVa, IVb from top to bottom, so that IVa and IVb represent the top and bottom halves of chamber IV respectively. The layout of the lest with section labels is plotted in Figure \ref{fig:SimNestSpread}. Using this notation, the nest exit is located in the bottom left corner of section IVb.

\begin{figure}
\centering
\begin{subfigure}{.5\textwidth}
  \centering
  \includegraphics[width=.9\linewidth, page=1]{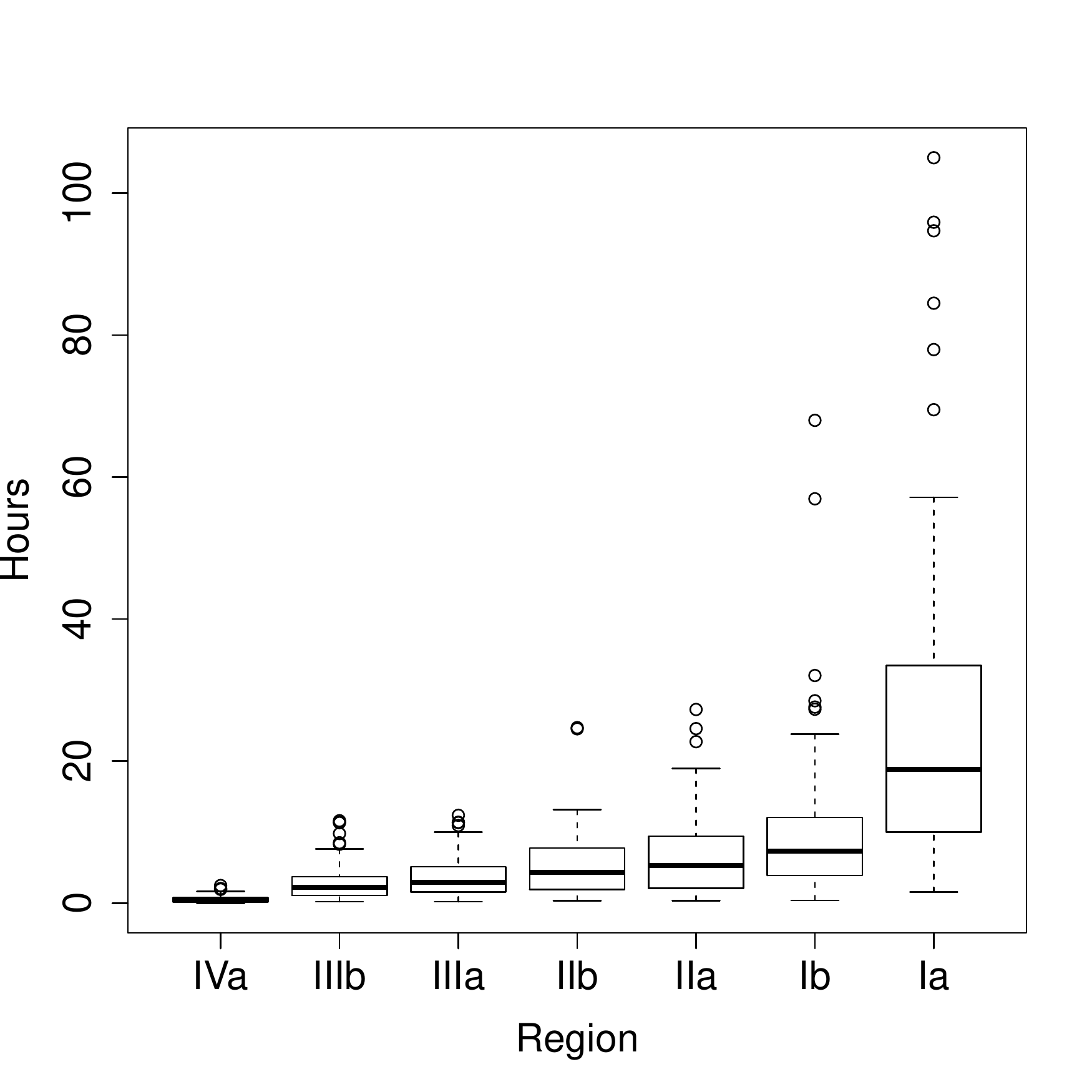}
  \caption{Time Until First Entry}
\end{subfigure}%
\begin{subfigure}{.5\textwidth}
  \centering
  \includegraphics[width=.9\linewidth, page=2]{boxplots}
  \caption{Time To Pass Through}
\end{subfigure}
\caption{Simulated Regional Spread}
\label{fig:Boxplots}
\end{figure}

 The first entry time of each of the 100 simulated ants into each of these segments was recorded. Boxplots of the results are shown in Figure \ref{fig:Boxplots}.

In Figure \ref{fig:Boxplots} (a), the total time in hours until entry into the designated region is plotted. The average time it takes to reach the top half of Chamber I, the farthest section from the next exit, is approximately 24 hours. As expected, the time to enter each of the chambers increases as you move up in the nest from chamber IVb through Ia. This is due to the fact that all ants are simulated to start at the nest doorway in chamber IVb. In Figure \ref{fig:Boxplots} (b), the time from entry in the previous region until entry into the designated region is plotted. This gives an impression of the amount of time it would take an agent to spread through each region of the nest. The simulation results indicate that on average, agents spread much more slowly through chambers I and IV than through chambers II and III. For example, on average it takes about 2.5 hours from entry into the lower half of chamber IV until entry into the top of chamber III, and about 14 hours on average to pass from the top of chamber I to the bottom of chamber I, making these the two regions that takes the longest to pass through. It only takes about 1 hour on average to pass from entry into the top half of chamber III until entry into the bottom half of chamber III. This is mainly due to the corridors of fast movement in the middle area of the nest captured by the motility surface in our model. The existence of a separate slower moving region at the entrance of the nest (chamber IV) thus may act to delay the rate at which an ant carrying an agent could spread deeper into the nest.  To gain a better understanding of the rate at which resources move through the nest, several model assumptions could be adjusted. Using a hierarchical model to allow for variability in individual movement behavior, for example, may better reflect the behavior of the individual ants. Additionally, time varying behavior is evident in the videos of ant movement. Incorporating this temporally varying behavior via latent discrete states \cite{langrock2012flexible} would impact the rate of spread. Further expanding the simulation to allow the agent to spread between interacting ants would also provide a more realistic model. Further research is needed to gain a better understanding of the relationship between animal movement behavior and resource spread dynamics.

\end{document}